\documentclass[twocolumn,showpacs,preprintnumbers, amsmath,amssymb,amsfonts, final, a4paper,aps, citeautoscript,footinbib,prb,superscriptaddress]{revtex4-1}

\usepackage{times}
\usepackage{graphicx}
\usepackage[latin1]{inputenc}
\usepackage{mathrsfs}
\usepackage{multirow}
\usepackage{epstopdf}

\usepackage[sort&compress]{natbib}

\usepackage[breaklinks=true,colorlinks=true,bookmarks=true,citecolor=blue,linkcolor=red,urlcolor=blue]{hyperref}
\graphicspath{{Figures/},{Figs/}}

\begin{document}
\title{Quantum phase transitions in multileg spin ladders with ring exchange}
\author{S.\ Capponi} 
\affiliation{Laboratoire de Physique Th\'eorique, CNRS UMR 5152,
  Universit\'e Paul Sabatier, F-31062 Toulouse, France.}
\author{P. Lecheminant}\email{philippe.lecheminant@u-cergy.fr}
\affiliation{Laboratoire de Physique Th\'eorique et
Mod\'elisation, CNRS UMR 8089,
Universit\'e de Cergy-Pontoise, Site de Saint-Martin,
2 avenue Adolphe Chauvin,
95302 Cergy-Pontoise Cedex, France.}
\author{M. Moliner} 
\affiliation{Laboratoire de Physique Th\'eorique et
Mod\'elisation, CNRS UMR 8089,
Universit\'e de Cergy-Pontoise, Site de Saint-Martin,
2 avenue Adolphe Chauvin,
95302 Cergy-Pontoise Cedex, France.}
\affiliation{IPCMS (UMR 7504) and ISIS (UMR 7006), Universit\'e de Strasbourg et CNRS, Strasbourg, France.}

\date{\today}

\begin{abstract}
Four-spin exchange interaction has been raising intriguing questions regarding the 
exotic phase transitions it induces in two-dimensional quantum spin systems. 
In this context, we investigate the effects of a cyclic four-spin exchange in the quasi-1D limit by 
considering a general $N$-leg spin ladder.
We show by means of a low-energy approach that, depending on its sign, this ring exchange interaction can engender 
either a staggered or a uniform dimerization from the conventional phases of spin ladders.
The resulting quantum phase transition is found to be described by the 
SU(2)$_N$ conformal field theory.
This result, as well as the fractional value of the central charge at the transition, is further confirmed by a large-scale numerical study performed by means of Exact Diagonalization and Density Matrix Renormalization Group approaches for $N \le 4$.

\end{abstract}

\maketitle   

\section{Introduction}

The destruction of Néel magnetic ordering at zero temperature by quantum 
fluctuations has attracted much interest in the past two decades. \cite{sachdev}
On top of the possible formation of spin-liquid phases,
strong quantum fluctuations in spin$-1/2$ systems might lead to unconventional
emerging quantum criticality. In this respect,
it has been proposed that there is, in two dimensions, a direct continuous quantum phase transition between 
the Néel antiferromagnetic phase and a valence bond solid (VBS) phase which breaks
spontaneously the translation symmetry. \cite{senthil_1, *senthil_2}
Such transition is beyond the conventional Ginzburg-Landau-Wilson paradigm since
direct transitions between phases whom order parameters enjoy different symmetries are 
generically first-order.
The exotic quantum critical point is described in terms of fractionalized quantities that are confined in the 
Néel and VBS phases and become deconfined at the transition. 
The resulting exotic quantum phase point has thus been dubbed
deconfined quantum criticality. \cite{senthil_1, *senthil_2}
The existence of this new class of quantum critical points has been extensively investigated numerically in
several quantum spin models in two dimensions 
\cite{sandvik_1, *sandvik_2,melko,kuklov_1, *kuklov_2,lou,Albuquerque2011,farnell,Nishiyama_1,*Nishiyama_2,zhao,KaulSUN,Ganesh}.
Most of these models contain four-spin interactions on top of the usual spin-exchange interaction.
For instance, the most studied model is the so-called $J-Q$ model with a special 
four-spin exchange interaction which is free from the sign problem. \cite{sandvik_1, *sandvik_2}
Quantum Monte Carlo computations in this case support the existence of a deconfined quantum critical
point between the Néel and VBS phases. \cite{sandvik_1, *sandvik_2}
However, the situation is still controversial since other studies suggest a weak first-order phase transition
which occurs for larger system sizes. \cite{kuklov_1, *kuklov_2}

\begin{figure}[ht]
\centering
\includegraphics[width=0.85\columnwidth,clip]{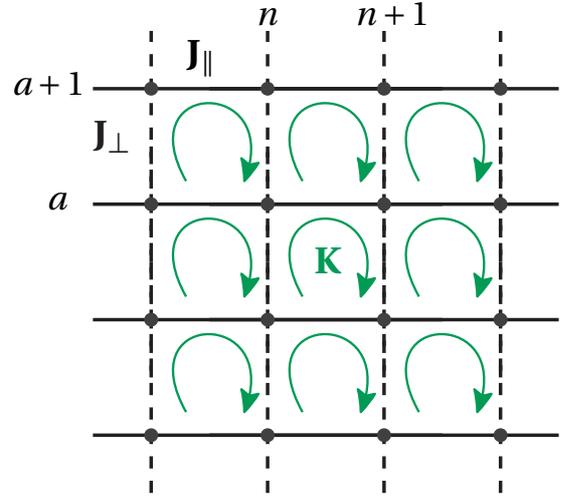}
\caption{(Color online) $N$-leg spin ladder with leg coupling $J_{\parallel}$, rung coupling $J_{\perp}$ and a ring-exchange interaction $K$.}
\label{fig:ladder}
\end{figure}

To get more insight into this problem, 
we investigate, in this paper, the nature of quantum phase transitions driven by four-spin exchange interactions 
in the quasi-1D limit where powerful techniques can be used. 
In this respect, we consider an $N$-leg spin ladder with a ring-exchange interaction with Hamiltonian:
\begin{eqnarray}
{\cal H} &=&  J_{\parallel} \sum_{a=1}^{N}\sum_n {\bf S}_{a,n} \cdot {\bf S}_{a,n+1} 
            + J_{\perp} \sum_{a=1}^{N-1} \sum_n {\bf S}_{a,n} \cdot {\bf S}_{a+1,n} \nonumber \\
           &+& K \sum_{\rm{plaquettes}}\left(P_{4} + P^{-1}_{4}  \right) ,
 \label{nlegringexchange}
\end{eqnarray}
where ${\bf S}_{a,n}$ is a spin$-1/2$ operator at the n$^{th}$ site of the ladder with leg index $a=1, \ldots, N$.
The parameters $J_{\parallel} >0$ and $ J_{\perp}$ are respectively 
the intrachain and the interchain exchange couplings.
The ring exchange operator $P_4$ is defined on each 
plaquette $\big[{\bf S}_{a, n},{\bf S}_{a,n+1},{\bf S}_{a+1,n+1},{\bf S}_{a+1,n}\big]$ between two consecutive chains and it cyclically permutes the states of the four spins on the plaquette (see Fig.~\ref{fig:ladder} and Eq.~(\ref{eq:ring-operator})).

This interaction appears in higher-order corrections in the strong-coupling expansion of the half-filled Hubbard model 
\cite{takahashi,girvin}
and is known to play a significant role in $^{3}$He adsorbed on graphite\cite{Roger-H-D-83,Fukuyama-08}.
The relevance of the four-spin cyclic exchange has also been reported 
\cite{Coldea01,Brehmer99,Matsuda00,Nunner02,Calzado03,Schmidt05,Bordas05,Notbohm07,Lake09} 
in the frame of inelastic neutron-scattering experiments for cuprates such as La$_2$CuO$_4$, La$_6$Ca$_8$Cu$_{24}$O$_{41}$,
La$_4$Ca$_{10}$Cu$_{24}$O$_{41}$, CaCu$_2$O$_3$, and SrCu$_2$O$_3$. 

The model (\ref{nlegringexchange}) in the $N=2$ case has been studied extensively over the years
by means of different analytical and numerical approaches\cite{Muller02,Lauchli03,Hikihara03_1,Hikihara03_2,uhrig,Hijii2002, Hijii2003,gritsev, Lecheminant05_1,Lecheminant05_2,Lin,hakobyan,hatsugai_1, hatsugai_2, nishimoto,totsuka}.
The zero-temperature phase diagram is rich and several exotic phases have been 
identified such as a scalar chirality phase which spontaneously breaks the time-reversal symmetry \cite{Lauchli03}.
When  $J_{\perp} >0$,  the ring exchange destabilizes the well known rung-singlet (RS) phase 
of the standard two-leg spin ladder and a staggered dimerization (SD) phase emerges.
The resulting quantum phase transition was predicted to be continuous and to belong to the 
SU(2)$_2$ universality class with central charge $c=3/2$ \cite{shura,gritsev}.
The latter result was confirmed numerically by means of Exact Diagonalization (ED) and
Density-Matrix Renormalization Group (DMRG) techniques \cite{Hijii2002, Hijii2003}.

In the general $N$ case, we are not aware, to the best of our knowledge, of any
investigation of the phase diagram of model (\ref{nlegringexchange}).
In the absence of the ring-exchange, it is well known that  the physics of the $N$-leg spin ladder
strongly depends on the parity of $N$ \cite{bookboso,giamarchi} . When $N$ is odd, 
a gapless phase with central charge $c=1$ is stabilized for all signs of $J_{\perp}$
while a RS phase and a non-degenerate gapful phase, which is equivalent to the Haldane phase
of the Heisenberg chain with spin $S=N/2$, appear in the $N$ even case  
respectively for $J_{\perp} >0$ and $J_{\perp} < 0$  \cite{bookboso,giamarchi}. 
As it will be shown in this work, the main effect of a weak  four-spin exchange interaction is to 
introduce dimerized phases (staggered or uniform dimerization depending on the sign of $K$) on top of these phases.
A central question is then the nature of the quantum phase transition driven by the ring-exchange interaction
for $N>2$. Can an exotic transition, like in the $N=2$ case, be stabilized?
The $N$ odd case is intriguing since, on general grounds, one expects a Berezinski-Kosterlitz-Thouless (BKT) 
quantum phase transition between a standard $c=1$ gapless phase to a dimerized one
as in the $J_1-J_2$ spin$-1/2$ Heisenberg chain \cite{haldanedimer}. In this respect,
is it possible to stabilize a more exotic transition with extended quantum criticality stemming from
the special character of the four-spin exchange interaction?
Finally, in relation with the deconfined quantum criticality paradigm, one may pose the question of the nature
of the quantum phase transition when $N \rightarrow \infty$ to reach the two-dimensional regime.

 In this paper, we investigate these questions by means of a
 low-energy approach when the chains are weakly-coupled and ED and
 DMRG computations for moderate couplings. We reveal the emergence of
 an exotic quantum phase transition in the SU(2)$_N$ universality
 class with fractional central charge $c = 3N/(N+2)$ for $N \le 4$.
 On the other hand, we expect, from the low-energy approach, a weak
 first-order transition for higher $N$.  The SU(2)$_N$ universality
 class has already been found in multicritical points of Heisenberg
 spin chains with polynomial interactions \cite{affleckhaldane}, and
 more recently in a spin chain with three-site interaction
 \cite{mila}. Similar quantum criticality appears also in the context
 of quantum Bose gases with long-range interactions
 \cite{heloise,tsvelikcoldatoms}. The SU(2)$_N$ criticality has also been shown to 
 describe Haldane-Shastry type spin chains with longer range interactions \cite{SierraWZW,ThomaleWZW}.

The rest of the paper is organized as follows.
The low-energy approach of model (\ref{nlegringexchange}) in the general $N$ case is presented in 
Sec.~\ref{sec:lowEnergy}. The emergence of the SU(2)$_N$ universality class is obtained through two different
methods: a semiclassical analysis and a more quantum approach based on the bosonization 
technique \cite{bookboso,giamarchi}. In Sec.~\ref{sec:DMRG}, we investigate numerically the quantum phase
transition for $N=2,3,4$ by means of ED and DMRG \cite{DMRG_1, *DMRG_2, *DMRG_3} computations.
Finally, our concluding remarks are given in Sec.~\ref{sec:Concluding} and some technical details on the continuum
limit are presented in the Appendix.
 
\section{Low-energy approach} 
\label{sec:lowEnergy}

In this section, we describe the low-energy approach of model (\ref{nlegringexchange}) 
in the weak-coupling limit for general $N$.
We assume that the in-chain exchange interaction $J_{\parallel}$ is antiferromagnetic
$J_{\parallel} >0$ and the chains are weakly coupled: $|J_{\perp},\, K |\ll J_{\parallel}$.

\subsection{Continuum limit}

In the absence of the interchain coupling ($J_{\perp} = 0$) and ring-exchange interaction ($K=0$), 
the model boils down to $N$ decoupled spin$-1/2$ SU(2) Heisenberg chains. As it is well known,
the latter model displays a quantum critical behavior in the SU(2)$_1$ universality
class with central charge $c= 1$ or equivalently with one bosonic gapless mode. \cite{bookboso}
 Using the non-Abelian bosonization approach, one can have a continuum description of the 
 lattice spin$-1/2$ operators in terms of their uniform and staggered parts \cite{bookboso}:
 \begin{equation}
a^{-1}_0 {\vec S}_{a,n} \simeq {\vec J}_{a } + (-1)^{x/a_0} {\vec n}_a,
\label{spinop}
\end{equation}
where $x= n a_0$ ($a_0$ being
the lattice spacing). 
The uniform part,  ${\vec J}_{a }  = {\vec J}_{a L}+ {\vec J}_{a R} $, expresses
in terms of  the left and right SU(2)$_1$ currents ${\vec J}_{a L,R} $ which generate
the underlying conformal symmetry of the spin$-1/2$ SU(2) Heisenberg chains. \cite{affleck86,affleckhaldane,bookboso}
In Eq.~(\ref{spinop}), the staggered magnetization  density of the $a^{th}$ chain takes the form:
\begin{equation}
\vec n_a = - \frac{ i\lambda}{ \pi a_0 \sqrt 2} {\rm Tr}\left( g_a {\vec \sigma} \right),
\label{staggop}
\end{equation}
where $\lambda$ is a non-universal real constant,  
 whose value has been obtained in Refs.~\onlinecite{affleck98,hikihara1998,lukyanov}, 
and ${\vec \sigma}$ stands for the Pauli matrices.
In Eq.~(\ref{staggop}), $g_a$ ($a=1,\ldots, N$) is the SU(2)$_1$ Wess-Zumino-Novikov-Witten (WZNW) primary field.
This field transforms under the spin$-1/2$ representation of SU(2) and has scaling dimension $\Delta = 1/2$.
\cite{bookboso,knizhnik,dms}
It is useful to recall that ${\rm Tr}\left( g_a {\vec \sigma} \right)$ is anti-Hermitian since
$g_a$ is an SU(2) matrix which explains the $i$ factor in Eq.~(\ref{staggop}).
In this continuum description, the non-interacting Heisenberg Hamiltonian of each chain expresses in terms of the 
currents:
\begin{eqnarray}
{\cal H}_0 &=& \frac{2\pi v}{3}  \sum_{a=1}^{N} \int dx \;  \left(
{\bf J}_{a L}^2  + {\bf J}_{a R}^2\right) 
 \nonumber \\
&-& a_0 \gamma \sum_{a=1}^{N} \int dx \; {\bf J}_{a L}\cdot {\bf J}_{a R}, 
\label{Hamnonint}
\end{eqnarray}
where $v=\pi J_{\parallel} a_0/2$ is the spin velocity
and $\gamma > 0$. 
The current-current interaction is thus a marginally irrelevant
contribution which gives rise to the 
well known logarithmic corrections
of the SU(2)$_1$ quantum criticality of the 
antiferromagnetic spin$-1/2$ Heisenberg chain. \cite{Affleck1989,shankar}
The last important operator in the continuum limit of model (\ref{nlegringexchange}) is the 
spin-dimerization field of the $a^{th}$ chain:
\begin{equation}
(-1)^n {\bf S}_{a,n} \cdot {\bf S}_{a,n+1} \sim  \epsilon_a =  \frac{\lambda}{\pi a_0 \sqrt 2} {\rm Tr}\left( g_a \right) ,
\label{dimerizationop}
\end{equation}
which has the same scaling dimension $\Delta=1/2$ as the staggered magnetization field. It is useful to observe, from
this representation, that the one-step translation symmetry $T_{a_0}$ corresponds
to a ${\mathbb{Z}}_2$  symmetry for the WZNW field: $g_a \rightarrow - g_a$.

With these results in hand, we can derive the continuum limit of the $N$-leg spin ladder with ring-exchange interaction.
The technical details are described in the Appendix for an $N$-leg spin ladder with  general 
four-spin exchange interactions. The interacting Hamiltonian  separates into two different parts:
 \begin{equation}
 {\cal H}_{int} =  {\cal H}_{\Delta = 1}  + {\cal H}_{cc} ,
 \label{conthamint}
 \end{equation}
where $ {\cal H}_{\Delta = 1}$ is the leading contribution with scaling dimension $\Delta =1$,
 which involves the staggered magnetization and dimerization fields:
 \begin{equation}
 {\cal H}_{\Delta = 1} =   \sum_{a=1}^{N-1} \int dx \; \left(  \lambda_1 {\bf n}_{a}\cdot {\bf n}_{a+1} 
 +  \lambda_2 \; {\epsilon}_{a} {\epsilon}_{a+1}  \right) ,
 \label{hamleading}
 \end{equation}
with $ \lambda_1 = a_0 J_{\perp}$ and $ \lambda_2=  36 a_0 K/\pi^2$ in the
case of a ring-exchange interaction (see the Appendix).
The second part of Eq.~(\ref{conthamint}) is marginal and expresses in terms 
of the currents:
\begin{eqnarray}
 {\cal H}_{cc} &=&   
  \lambda_3 \sum_{a=1}^{N} \int dx \; 
   {\bf J}_{aL} \cdot {\bf J}_{aR} 
\nonumber \\   
 &+&  \lambda_4 \sum_{a=1}^{N-1} \int dx \; 
 \left( {\bf J}_{a L} \cdot {\bf J}_{a+1 R} 
+ {\bf J}_{a R} \cdot {\bf J}_{a+1 L} \right) ,
 \label{hamcurrent}
 \end{eqnarray}
 where we have neglected all chiral interactions 
and thus velocity renormalization terms. The identification of the coupling constants
is detailed in the Appendix and we find:
$ \lambda_3 = a_0 \left[ - \gamma + 
2 K \left( 1 - \lambda^2 \right)
+ 4 \lambda^2 K  \left(
3 \lambda^2 
- 1  \right)/ \pi^2  \right] $ 
and $ \lambda_4 = a_0 \left(J_{\perp} + 4 K - 4K (1+ 4\lambda^4)/\pi^2 \right)$.
It is important to observe that a $N$-leg spin ladder with more general four-spin exchange
interaction will have the same continuum description in the weak-coupling limit. 
The actual form of the four-spin exchange interaction is encoded in the expression of the coupling
constants  $\lambda_{1,2,3,4}$ (see the Appendix). This means that our conclusion on the quantum phase
transition for the spin ladder (\ref{nlegringexchange}) still holds for a more general interaction.
In particular, the effective Hamiltonian (\ref{hamleading}) is the most relevant perturbation
for general translation-invariant spin-1/2 ladders with SU(2) symmetry. As we will see, the competition
between the two terms in Eq. (\ref{hamleading}) is responsible for the emergence of an exotic quantum phase
transition.

To this end, the strategy is to (i) analyze the effects of ${\cal H}_{\Delta = 1}$, 
which is a strongly relevant perturbation,  (ii) determine the different phases
of the model in the weak-coupling regime 
and (iii) discuss the nature of the quantum phase transition. The next step
is then to investigate the stability of these
results with respect to the marginal current-current interactions (\ref{hamcurrent}).

\subsection{Semiclassical approach}

We will first consider a naive  semiclassical approach of model (\ref{hamleading}) 
which gives some hints about the possible phases  and transitions of the problem.
A more quantum approach will be described in the next subsection to further justify 
the results obtained within the semiclassical approach.

Let us denote $W(g)$ the action of the SU(2)$_1$ WZNW model which 
describes the SU(2)$_1$ criticality of the spin$-1/2$ SU(2) Heisenberg chain.
This action reads as follows \cite{knizhnik,dms,witten}:
\begin{eqnarray}
W(g) &=& \frac{1}{8\pi} \int d^2 x \; {\rm Tr}(\partial^{\mu} g^{\dagger} \partial_{\mu} g) 
+ \Gamma(g), \nonumber \\
\Gamma(g) &=& \frac{-i}{12\pi}   \int_B d^3 y \; \epsilon^{\alpha \beta \gamma} 
 {\rm Tr}(g^{\dagger} \partial_{\alpha} g g^{\dagger} \partial_{\beta} g g^{\dagger} \partial_{\gamma} g),
\label{WZNW}
\end{eqnarray}
where $\Gamma(g)$ is the famous WZNW topological term.
The effective action of model (\ref{hamleading}) can be obtained 
using the definitions (\ref{staggop}, \ref{dimerizationop}). 
In this respect, it is convenient to write the product of staggered magnetization operators as
follows:
 \begin{eqnarray}
 {\vec n}_a  \cdot  {\vec n}_{a+1}  &=&  -  \frac{\lambda^2}{2 \pi^2 a_0^2} 
 {\rm Tr}\left( g_a {\vec \sigma} \right) {\rm Tr}\left( g_{a+1} {\vec \sigma} \right) 
 \nonumber \\
 &=&
  \frac{\lambda^2}{2\pi^2 a_0^2} 
 {\rm Tr}\left( g_a {\vec \sigma} \right) {\rm Tr}\left( g^{\dagger}_{a+1} {\vec \sigma} \right),
\label{staggsub}
\end{eqnarray}
where we used the fact that ${\rm Tr}\left( g_{a+1} {\vec \sigma} \right) $ is anti-Hermitian
for an SU(2) matrix. Using the completeness relation
$ \sigma^{i}_{\alpha \beta}  \sigma^{i}_{\gamma \delta} = 2 ( \delta_{\alpha \delta} 
 \delta_{\beta \gamma} -   \delta_{\alpha \beta}   \delta_{\gamma \delta}/2 )$,
 the action of model (\ref{hamleading}) then takes the form:
\begin{equation}
{\cal S} =   \sum_{a=1}^{N} W(g_a) + \int d^2 x \;  (V_1 + V_2) , 
\label{Stot}
\end{equation}
with
\begin{eqnarray}
V_1 &=&      \frac{\lambda_1  \lambda^2}{\pi^2 a_0^2}  \sum_{a=1}^{N-1} {\rm Tr}( g_a  g^{\dagger}_{a+1}) 
\nonumber \\
V_2 &=& ( - \lambda_1 +  \lambda_2) \frac{\lambda^2}{2 \pi^2 a_0^2}  \sum_{a=1}^{N-1} {\rm Tr}( g_a )  {\rm Tr}(g_{a+1}) .
\label{Vleg}
\end{eqnarray}
The two contributions of Eq.~(\ref{Vleg}) are of different nature.
In particular, $V_1$ is invariant under an SU(2)$_L$ $\times$ SU(2)$_R$ symmetry: 
$g_a  \rightarrow U g_a V$, $U$ and $V$ being independent SU(2) matrices.
In contrast, $V_2$ is only SU(2) invariant: $g_a  \rightarrow U g_a U^{\dagger}$.

We will now apply a semiclassical approach to investigate the nature
of the zero-temperature phases of model (\ref{Stot}). 
$V_1$ is a strongly relevant perturbation with scaling dimension $\Delta = 1$ and a
gap will open for some degrees of freedom. The main hypothesis of the semiclassical approach
is to assume that $V_1$ operator gives the largest gap $\Delta_1$ of the problem.
This hypothesis is expected to be valid in some parts of the phase diagram of the lattice 
model  (\ref{nlegringexchange}). The numerical simulations of Sec.~\ref{sec:DMRG} will shed light on 
the correctness of this semiclassical approach. Once this assumption has been made,
the next step is then to write down an effective action which captures the low-energy
properties of the model when $E \ll \Delta_1$. The nature of the effective action
turns out to depend on the sign of $ \lambda_1 = a_0 J_{\perp}$. Next we discuss the two cases $J_{\perp} > 0$ and $J_{\perp} < 0$.

\subsubsection{$J_{\perp} >0$}
Since $ \lambda_1 >0$, the configuration $g_{a+1} = - g_a$ ($a=1, \ldots, N-1$) minimizes the 
potential $V_1$ (\ref{Vleg}) over SU(2) matrices. 
Averaging out the $g_{2,\ldots, N}$ fields, we obtain an effective action on the $g_1$ field:
\begin{equation}
{\cal S}_{eff} =  N W(g_1) - N ( - \lambda_1 +  \lambda_2)  \frac{\lambda^2}{2 \pi^2 a_0^2}   \int d^2 x \;  
 \left({\rm Tr}(g_1) \right)^2    ,
\label{actiongeffnlegAF}
\end{equation}
where we used $W(-g) = W(g)$.
The latter model is the SU(2)$_N$ WZNW model perturbed by $\left({\rm Tr}(g_1) \right)^2 \equiv
{\rm Tr}(\Phi^{(1)}) $, $\Phi^{(1)}$ being the spin$-1$ WZNW  primary field with scaling dimension $4/(N+2)$.
This model corresponds to the effective action for the spin$-N/2$ Heisenberg model obtained
by Affleck and Haldane. \cite{affleckhaldane}
The IR property of model (\ref{actiongeffnlegAF}) depends on the sign of $\lambda_1 -  \lambda_2$ and
the parity of $N$. If $\lambda_1 -  \lambda_2 > 0$, i.e. for a weak four-spin exchange interaction,
the minimum condition on the potential of the action (\ref{actiongeffnlegAF})  corresponds
to an SU(2) matrix with the constraint: ${\rm Tr}(g_1) = 0$. Since for an SU(2) matrix,
we have the decomposition: $g_1 = n_0 I + i {\vec \sigma} \cdot  {\vec n} $ with $n_0^2 + {\vec n}^2 =1$,
the constraint ${\rm Tr}(g_1) = 0$ gives: $g_1 = i {\vec \sigma} \cdot  {\vec n} $, ${\vec n} $ being an unit vector.
Plugging this expression into Eq.~(\ref{WZNW}), one obtains the non-linear $\sigma$ model
with a topological term $\theta = N \pi$.  \cite{affleckhaldane}
When $N$ is even, we have a massive non-degenerate phase which corresponds to 
the well known RS phase of the $N$-leg spin ladder for $J_{\perp} >0$. When $N$ is odd, the  non-linear $\sigma$ model with 
topological term $\theta =\pi$ is known to be gapless in the SU(2)$_1$ universality class.
\cite{shankaread,zamolos} 
One recovers the standard gapless phase of the spin ladder with an odd number of legs.

When $\lambda_1 -  \lambda_2 < 0$, i.e. $K > \pi^2 J_{\perp} /36$, 
the minimum condition on the potential of the action (\ref{actiongeffnlegAF}) 
is $g_1 = \pm I$ for all $N$. One enters a dimerized phase $\langle {\rm Tr}(g_a) \rangle \ne 0$
which is two-fold degenerate and breaks spontaneously the one-step translation symmetry
$T_{a_0}$ ($g_a \rightarrow - g_a$). 
The dimerization pattern in a given chain is out-of-phase with the one in the neighboring chains: $\langle {\rm Tr}(g_1) \rangle = - \langle {\rm Tr}(g_2) \rangle = \ldots =
(-1)^{N-1} \langle {\rm Tr}(g_N) \rangle $ so that $\langle (-1)^n {\bf S}_{1,n} \cdot {\bf S}_{1,n+1} \rangle = 
- \langle(-1)^n {\bf S}_{2,n} \cdot {\bf S}_{2,n+1} \rangle = \ldots =  (-1)^{N-1} \langle (-1)^n {\bf S}_{N,n} \cdot {\bf S}_{N,n+1} \rangle \ne 0$.
 This phase corresponds to the SD phase which exists for all $N$.

The location of the quantum phase transition between the RS phase when $N$ is even (or the gapless phase when $N$ is odd), 
and the SD phase does not depend on $N$. It is located at $\lambda_1 = \lambda_2$,
i.e. $J_{\perp} = 36 K/\pi^2$.
The action that controls the transition is ${\cal S}_{eff} =  N W(g_1)$
and takes then the form of the SU(2)$_N$ WZNW model.
We thus expect a quantum phase transition that belongs to the SU(2)$_N$ 
universality class with fractional central charge $c= 3N/(N+2)$.
The leading asymptotics of the spin-spin correlation functions at the quantum critical point
can be estimated since $g_1$ is the spin$-1/2$ SU(2)$_N$ primary field with
scaling dimension $3/2(N+2)$ \cite{dms}. The equal-time spin-spin correlation reads then as follows:
\begin{equation}
\langle  {\bf S}_{a,n}    {\bf S}_{a,n+m}  \rangle \sim \frac{(-1)^{m} (\ln m)^{1/2}}{m^{3/(N+2)}} ,
\label{correl}
\end{equation}
where the logarithmic corrections are the same as for spin$-1/2$ \cite{Affleck1989}.
Similarly, the dimer-dimer correlation function is also algebraic with the same power-law behavior but
different  logarithmic corrections:
\begin{equation}
\langle  {\epsilon}_{a} (x)    {\epsilon}_{a} (0)  \rangle \sim \frac{ (\ln x)^{-3/2}}{x^{3/(N+2)}} .
\label{correldimer}
\end{equation}
When $N$ is odd, the quantum critical point has additional gapless modes with respect to the  standard gapless
Heisenberg chain criticality with central charge $c=1$. The power-law decay of the spin-spin correlation
function  at the SU(2)$_N$ critical point (\ref{correl}) is then different from the standard $1/x$ scaling of the
spin-1/2 Heisenberg chain.

For $N=2$, the semiclassical approach reproduces the  SU(2)$_2$  transition, first predicted
in Ref.~\onlinecite{shura} using an exact solution of the perturbation (\ref{hamleading}) 
based on the Majorana fermions formalism. The location of the transition, obtained
within our approach, is $K/J_\perp = \pi^2/36 \simeq 0.27415$.
Surprisingly enough, this estimate is in good agreement with a previous DMRG
analysis that found a closure of the spin gap for
$K/J_\perp \simeq 0.3$ with $J_\perp=1$ \cite{Hikihara03_1, *Hikihara03_2}.  A second DMRG study 
locates the transition in the window $0.2<K/J_\perp<0.26$ \cite{Lauchli03}.
We will come back to this question in our numerical investigation in Sec.~\ref{sec:DMRG}. 

\subsubsection{$J_{\perp} < 0$}

When $J_{\perp} < 0$, i.e. $\lambda_1 < 0$, the minimum condition for $V_1$ is now
$g_{a+1} = g_a$ ($a=1, \ldots, N-1$).
Averaging out the $g_{2,\ldots, N}$ fields, we obtain an effective action on the $g_1$ field
which describes the low-energy properties of the model when $E \ll \Delta_1$:
\begin{equation}
{\cal S}_{eff} =  N W(g_1) + N ( - \lambda_1 +  \lambda_2)   \frac{\lambda^2}{2 \pi^2 a_0^2}   \int d^2 x \; 
 \left({\rm Tr}(g_1) \right)^2    .
\label{actiongeffnlegF}
\end{equation}
When $-\lambda_1 +  \lambda_2 >0$, in the semiclassical approach, $g_1$ satisfies the constraint
${\rm Tr}(g_1) = 0$. The effective action (\ref{actiongeffnlegF}) becomes equivalent to the non-linear $\sigma$ model
with a topological term $\theta = N \pi$. When $N$ is even, it describes the Haldane phase
of the Heisenberg spin chain $S=N/2$ obtained when $ J_{\perp} \rightarrow - \infty$.
In contrast, when $N$ is odd, the phase is gapless as in the $ J_{\perp} > 0$ case.

For sufficiently large negative K, one enters in a new phase when $-\lambda_1 +  \lambda_2 < 0$ 
which is described by $g_1 = \pm I$ for all $N$. This phase corresponds to a uniform
dimerization (UD) phase where : 
$\langle {\rm Tr}(g_1) \rangle = \langle {\rm Tr}(g_2) \rangle = \ldots = \langle {\rm Tr}(g_N) \rangle$.  
This time, the dimerization in a given chain is in-phase with the one in the neighboring chains: $\langle (-1)^n {\bf S}_{1,n} \cdot {\bf S}_{1,n+1} \rangle = \langle (-1)^n {\bf S}_{2,n} \cdot {\bf S}_{2,n+1} \rangle = \ldots = \langle (-1)^n {\bf S}_{N,n} \cdot {\bf S}_{N,n+1} \rangle \ne 0$.
 There is thus a quantum phase transition between the Haldane phase (and respectively the gapless phase) and the UD phase when $N$ is even (respectively odd).
The nature of the  transition at $\lambda_1 = \lambda_2$  for model (\ref{nlegringexchange}) is 
the same as in the $J_{\perp} > 0$ case and belongs to the SU(2)$_N$ 
universality class. The leading asymptotics of the correlation functions at the transition are also
given by Eqs. (\ref{correl}, \ref{correldimer}).

\subsection{Abelian bosonization approach}

Using a simple semiclassical approach, we have seen that
an exotic  quantum phase transition in the   SU(2)$_N$ 
universality class might emerge in model (\ref{nlegringexchange}) for all $N$.
Here, we present an alternative approach, based on the Abelian bosonization,
which confirms the semiclassical prediction.

 In the Abelian bosonization approach, the SU(2)$_1$ criticality of the decoupled
 spin$-1/2$ Heisenberg chain is described by a bosonic field. \cite{bookboso}
 Introducing $N$ bosonic field $\Phi_a$, the low-energy Hamiltonian in the absence
 of interactions is:
 \begin{equation}
{\cal H}_{0} =  \frac{v}{2} \sum_{a=1}^{N} \int dx \; \left( \left(\partial_x  \Phi_a \right)^2 + 
  \left(\partial_x  \Phi_a \right)^2 \right)  ,
\end{equation}
where $\Theta_a$ ($\Theta_a = \Phi_{aL} - \Phi_{aR}$) are dual fields,
$\Phi_{aL,R}$ being the chiral components of the Bose fields $\Phi_a = \Phi_{aL} + \Phi_{aR}$.
The staggered magnetization (\ref{staggop}) and the dimerization 
operator (\ref{dimerizationop}) can be expressed in terms of these bosons \cite{bookboso,shelton}: 
\begin{eqnarray}
{\bf n}_a &=& \frac{\lambda}{\pi a_0} \Big(\cos\big(\sqrt{2\pi}\;\Theta_a \big), \sin\big(\sqrt{2\pi}\;\Theta_a \big),
- \sin\big(\sqrt{2\pi}\;\Phi_a \big) \Big) 
\nonumber \\
\epsilon_a &=& \frac{\lambda}{\pi a_0} 
\cos\left(\sqrt{2\pi}\; \Phi_a \right) .
\label{boserep}
\end{eqnarray}
Using these results, the leading contribution (\ref{hamleading}) can be bosonized:
 \begin{eqnarray}
{\cal H} &=& 
  \frac{v}{2}  \sum_{a=1}^{N}  \int dx \; \left[  \left(\partial_x \Phi_a 
\right)^2 +  \left(\partial_x \Theta_a \right)^2 \right]
\nonumber \\
&+& \int dx \; \sum_{a=1}^{N-1} \left[ g_{\perp}
\cos \left( \sqrt{2\pi} \left( \Theta_{a+1} -  \Theta_{a} \right) \right) \right.
\nonumber \\
&+& \left.  g_{\perp} \sin \left( \sqrt{2\pi} \Phi_{a+1} \right) 
 \sin \left( \sqrt{2\pi} \Phi_{a} \right) 
  \right.
\nonumber \\
&+& \left.  g_4 \cos \left( \sqrt{2\pi} \Phi_{a+1} \right)  \cos \left( \sqrt{2\pi} \Phi_{a} \right)  \right] ,
\label{hameff}
\end{eqnarray}
where $g_{\perp}  = J_{\perp}\lambda^2 /a_0 \pi^{2}$ 
and $g_4 = 36 K \lambda^2 /a_0\pi^{4} $. This model has an interesting symmetry content 
which is explicit within the bosonization formalism. Indeed,
one can observe that model (\ref{hameff}) is invariant under the Gaussian duality 
symmetry $ \Phi_{a} \leftrightarrow \Theta_a$
by fine-tuning the interaction $g_{\perp} = g_4$:
\begin{eqnarray} 
{\cal H}_{SD} &=&   
  \frac{v}{2}  \sum_{a=1}^{N}  \int dx \; \left[  \left(\partial_x \Phi_a 
\right)^2 +  \left(\partial_x \Theta_a \right)^2 \right]
\nonumber \\
&+ g_{\perp}& \int dx \; \sum_{a=1}^{N-1} \left[ 
\cos \left( \sqrt{2\pi} \left( \Theta_{a+1} -  \Theta_{a} \right) \right) \right.
\nonumber \\
&+& \left.  \cos \left( \sqrt{2\pi} \left( \Phi_{a+1} -  \Phi_{a} \right) \right)  \right] .
\label{sdsg} 
\end{eqnarray}
The fine-tuning $g_{\perp} = g_4$, i.e. $K/J_\perp = \pi^2/36$, corresponds
to the location of the quantum phase transition for all signs of $J_\perp$ obtained within the semiclassical approach. 
The low-energy effective theory (\ref{sdsg}), which enjoys a  self-dual symmetry, has been 
found in totally different contexts. On the one hand, the deconfining phase transition of 
the 2+1-dimensional SU($N$) Georgi-Glashow model is controlled by model  (\ref{sdsg}). \cite{phle}
On the other hand, as shown recently, it describes the quantum phase transition in
dipolar quantum Bose gas. \cite{heloise,tsvelikcoldatoms}
The model is also relevant to the competition between superconductivity and charge-density wave 
or between superfluidity and solid, or supersolids in quasi-dimensional
systems. \cite{tsvelik_1,fradkin_1,carrcoldatoms}

Though the interaction in model (\ref{sdsg}) is strongly relevant with scaling dimension $\Delta=1$, 
a quantum critical behavior, stemming from the special form of the interaction, is expected.
Indeed, model (\ref{sdsg}) is invariant under two independent global U(1) transformations:
\begin{eqnarray}
 \Phi_{a} &\rightarrow&  \Phi_{a} + \alpha \nonumber \\
 \Theta_{a} &\rightarrow&  \Theta_{a} + \beta ,
 \label{U1sym}
\end{eqnarray}
$\alpha, \beta$ being real numbers. The transformation (\ref{U1sym}) 
gives a U(1)$_L$ $\times$ U(1)$_R$ global continuous symmetry of model (\ref{sdsg}). One thus expects 
the model to be gapless with at least one bosonic field protected by the symmetry  (\ref{U1sym}).
 In fact, it has been shown that model (\ref{sdsg}) displays an extended SU(2)$_N$ quantum 
 critical behavior. \cite{phle,heloise}
One way to  establish this result is to use the following conformal embedding 
\begin{equation}
\text{SU(2)}_1 \times \text{SU(2)}_1 \times \cdots
\times \text{SU(2)}_1 \rightarrow \text{SU(2)}_N \times {\cal G}_N,
\label{embedgenn}
\end{equation}
where  ${\cal G}_N$ is a discrete conformal field theory (CFT) with
central charge $c_{{\cal G}_N}=N - 3N/(N+2) = N(N - 1)/(N+2)$.
It has been shown in Refs.~\onlinecite{phle,heloise} that the self-dual perturbation of model (\ref{sdsg})  
corresponds to a special primary field of the ${\cal G}_N$ CFT. 
A spectral gap opens  for the discrete ${\cal G}_N$ degrees of freedom, leaving the SU(2)$_N$ ones intact. 
We thus conclude that model (\ref{sdsg})  displays critical properties in the SU(2)$_N$ universality class.
The sign of the coupling constant $g_{\perp}$ does not play a crucial role for the emergence 
of this quantum criticality since one can freely change the sign of the perturbation  (\ref{sdsg}) 
by the transformation:
 \begin{eqnarray}
 \Phi_{2n}  &\rightarrow& \Phi_{2n} + \frac{\sqrt{\pi}}{2}, \;  \; \Phi_{2n+1}  \rightarrow \Phi_{2n+1}
 \nonumber \\
  \Theta_{2n}  &\rightarrow& \Theta_{2n} + \frac{\sqrt{\pi}}{2},  \;  \; \Theta_{2n+1}  \rightarrow \Theta_{2n+1} .
 \label{cantrans}
\end{eqnarray}

We thus confirm the conclusion of the semiclassical ana\-ly\-sis: interaction (\ref{hamleading}) 
exhibits an exotic quantum phase transition in the SU(2)$_N$ universality class
for all signs of $J_{\perp}$. 
The connection between the two approaches stems from the bosonization of 
the WZNW field $g_a$, which follows from the identification (\ref{boserep}):
\begin{eqnarray}
g_a  = \frac{1}{\sqrt{2}}
\left(
\begin{array}{lccr}
 e^{-i \sqrt{2\pi} {\Phi}_a}  &
i  e^{-i \sqrt{2\pi} {\Theta}_a}  \\
i  e^{i \sqrt{2\pi} {\Theta}_a}   &
 e^{i \sqrt{2\pi} {\Phi}_a} 
\end{array} \right) .
\label{gbosWZNW}
\end{eqnarray}
The self-dual sine Gordon perturbation of Eq.~(\ref{sdsg}) is then easily shown to be equal to 
$V_1$ (\ref{Vleg}) which governs the quantum phase
transition in the semiclassical approach. The main interest of the approach based on 
the conformal embedding (\ref{embedgenn}) is to show non-perturbatively that $V_1$ gives a mass
gap for the discrete degrees of freedom that is independent from the SU(2)$_N$ ones.

Though we have shown that the critical properties of model (\ref{hamleading}) are governed
by an SU(2)$_N$ CFT,  we are not guaranteed that the quantum phase transition of the initial model
  (\ref{nlegringexchange}) belongs to the SU(2)$_N$ universality class.
  Indeed, in general, such fixed points with extended criticality
 are fragile since they can be destabilized by several relevant primary operators. 
 In addition, the marginal current-current interaction (\ref{hamcurrent}), that we have neglected so far,
 might destroy the criticality of the transition into a first-order phase transition.
 The spectrum of the SU(2)$_N$ CFT is well known.\cite{dms} 
 The primary operators of SU(2)$_N$ CFT are labeled by their spins $j=1/2,1,\ldots,N/2$
 and have scaling dimension $ 2 j(j+1)/(N+2)$. 
As the consequence of the  SU(2) symmetry of the lattice model (\ref{nlegringexchange}), they take the form $({\rm Tr}(G))^m, m=2j$, $G$ being the SU(2)$_N$ WZNW field.
Furthermore, the translational invariance restricts $m$ to be even since $G \rightarrow -G$ under $T_{a_0}$.
The most relevant operator is thus the spin$-1$ operator, i.e. $({\rm Tr}(G))^2$,
which, according to the semiclassical approach, leads to the different phases
of the problem. This operator is canceled by the fine-tuning of the four-spin exchange interaction.
 The next relevant operator is the spin$-2$ primary field ($m=4$) with scaling dimension
 $12/(N+2)$ which is a strongly relevant perturbation when $N>4$. 
This operator is not forbidden by any symmetry of the original lattice.
 Though it requires a proof, it is 
 thus likely that the phase transition is first-order for $N>4$.
 In this respect, the quantum Monte-Carlo approach of Ref. \onlinecite{tsvelikcoldatoms} points out a 
 first-order phase transition in the $N$ identical tubes model with polar molecules when $N>4$.
 For $N \le 4$, the only possible source of  a mass gap is a
 marginal perturbation. In this respect, since the SU(2)$_N$ fixed point has been obtained
 for all sign of $J_{\perp}$, this criticality is then expected to be stable under the current-current interaction 
 at least when $J_{\perp} < 0$. In the $J_{\perp} > 0$ case, one may expect that this quantum criticality 
 still appears at least in the weak-coupling regime due to the dominance of the in-chain marginal irrelevant current-current
 interaction. Let us also notice that, as it is a marginal interaction that may drive the transition to first order,
 the first-order jump is expected to be very weak and the properties of the SU(2)$_N$ critical point might be seen
 in the vicinity of the transition. 
 We thus conclude that the quantum phase transition in the  SU(2)$_N$ universality class
 occurs for $N=2,3,4$ with respectively central charge $c=3/2,9/5,2$.
 Large-scale numerical simulations are clearly call for to reveal the emergence of this exotic quantum-critical
 behavior.
 
\section{Numerical simulations}\label{sec:DMRG}

In this section, we present extensive numerical simulations of the
$N$-leg spin$-1/2$ ladder with cyclic exchange model
(\ref{nlegringexchange}).  We take the antiferromagnetic leg coupling
$J_\parallel=1$ as the unit of energy.  Most data are obtained with
DMRG simulations with open boundary conditions (OBC). Note that, since
a SD phase can occur when $J_\perp>0$, we have added two extra sites
to some of the chains in order to prevent a ground-state
degeneracy.~\cite{Lauchli03} Typically, we have used 16 sweeps and
kept up to 1600 states which is sufficient for convergence,
i.e. having a discarded weight of at most $10^{-6}$.  When $K=0$, the
physics of these spin$-1/2$ ladders is well known~\cite{Dagotto1996}
and depends on the parity of $N$: for $N$ odd (respectively even) the
system is gapless (respectively gapped), for both signs of $J_\perp$.
When the ratio $K/J_\perp$ increases, our numerical simulations confirm
the existence of a quantum phase transition towards a dimerized state
that breaks translation symmetry and corresponds to a uniform
(respectively staggered) pattern between chains for $J_\perp$ and $K$
negative (respectively positive).

Let us mention that Exact Diagonalization (ED) using Lanczos technique
can also be used to locate the phase transition, as there is a clear
change of quantum numbers of the first excited state across it. This
so-called level spectroscopy technique has often been used in similar
context~\cite{Nomura1994,Okamoto1992,Schulz1986,Affleck1989}. 
There is also the possibility to investigate the level crossing using twisted boundary conditions (TBC), as was done for instance in Ref.~\onlinecite{Hijii2009} with a similar model. 
It turns out that, when available, the latter technique tends to reduce the finite-size effects in the location of the critical point.
Indeed, while periodic boundary conditions (PBC) engender discrete momenta, which can spoil the measure of momentum-dependent operators, adding a twist in the boundary conditions allows a continuous definition of the momenta and overcomes this problem.~\cite{Poilblanc_1991}

Next, we present our numerical results obtained for the values of $N$ for which the low-energy study Sec.~\ref{sec:lowEnergy} predicts a phase transition in the SU(2)$_N$ universality class when $N=2,3,4$. Both ferromagnetic and antiferromagnetic $J_{\perp}$ couplings will be considered. 

\subsection{Two-leg ladders ($N=2$)}\label{sec:N2}
\subsubsection{$N=2$ antiferromagnetic rung coupling case: $J_\perp=1.0$}\label{sec:N2J1}

\begin{figure}[ht]
\centering
\includegraphics[width=0.95\columnwidth,clip]{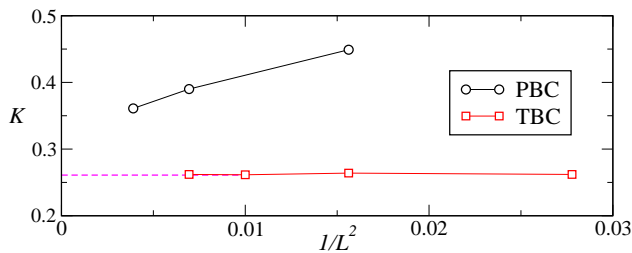}
\caption{(Color online) $N=2$ and $J_\perp=1$: crossing point with PBC (circles) and TBC (squares) versus system size (see text for definitions). The most precise extrapolations are obtained using TBC data and lead to a critical point at $K=0.26$ (dashed line).}
\label{fig:ED_N2_J1_AF}
\end{figure}

First, we use ED simulations with TBC in order to locate the critical
point.  Using quantum numbers to label respectively (i) the
reflection perpendicular to the rungs (ii) the reflection
perpendicular to the legs (i.e. that cuts the bonds across the boundary conditions) and (iii) the spin reversal symmetry, a simple analysis shows that RS and
SD phases are identified in sectors $(+1,+1,+1)$ $(+1,-1,-1)$
respectively.~\footnote{We only consider an even length $L$ in order to accomodate the SD phase. In this case, the RS phase is made of an even number of singlets and thus is even with respect to all of the three symmetries.}  The results for this level crossing are shown for
different system sizes in Fig.~\ref{fig:ED_N2_J1_AF}, which leads to
an estimate of the critical value at $K/J_\perp=0.26$. Note that for
comparison, we also plot the level crossing found by comparing the
lowest triplet and singlet states at momentum $\pi$ using PBC. Using
this criterion, we find a critical value $K/J_\perp=0.33$. However,
since it has stronger finite-size effects, we expect this result to be
less accurate. Our finding is slightly above the estimated $K=0.19$
found numerically in Refs.~\onlinecite{Hijii2002,Hijii2003}, but in
better agreement with $K=0.23\pm 0.03$ found in
Ref.~\onlinecite{Lauchli03}.  Surprisingly enough, the estimate found
in the weak-coupling approach ($K \simeq 0.27$) is very close to the
numerical result obtained for moderate coupling.

\begin{figure}[ht]
\centering
\includegraphics[width=0.95\columnwidth,clip]{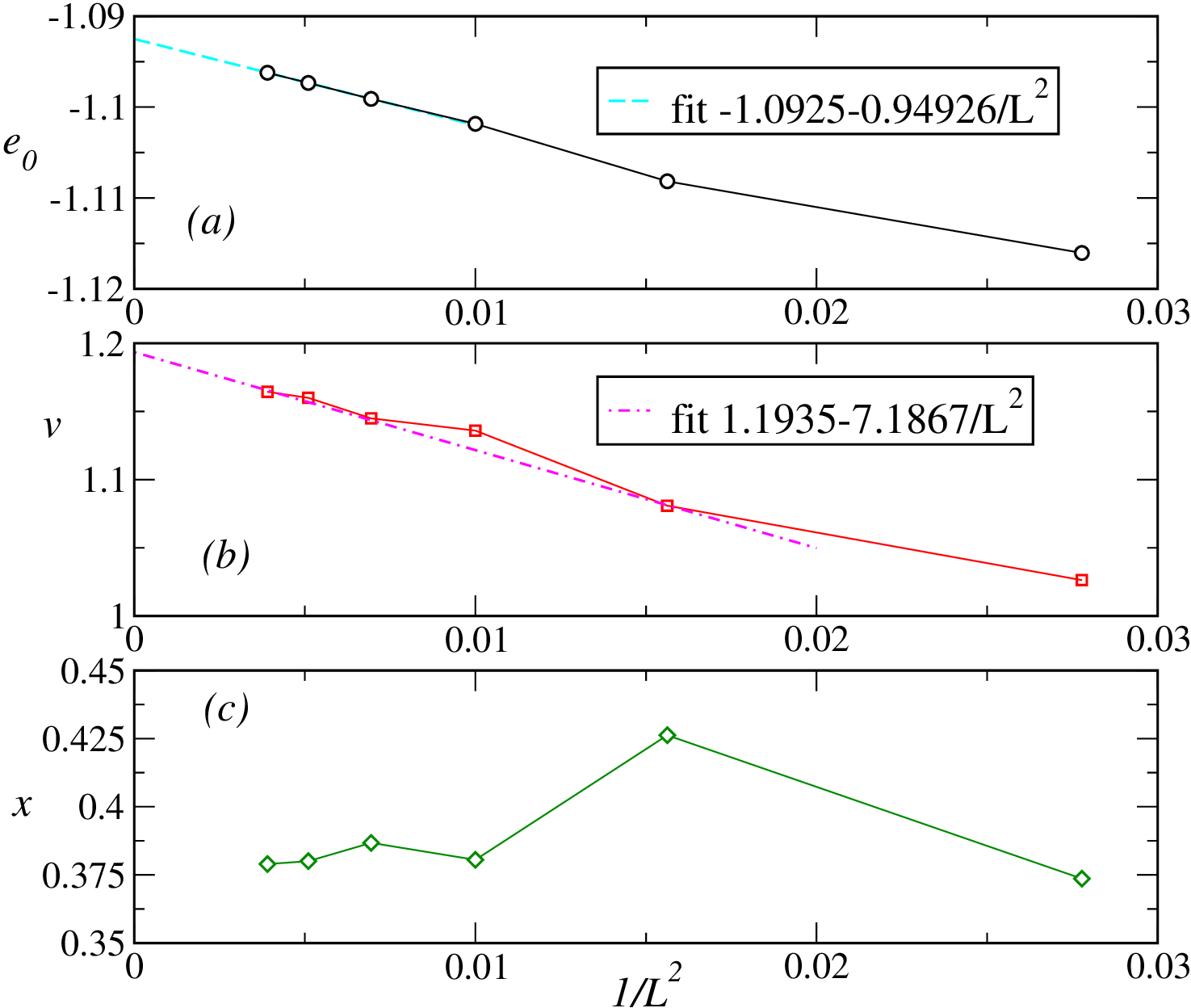}
\caption{(Color online) $N=2$ and $J_\perp=1$: finite-size scaling of various quantities at the critical point $K=0.26$: 
(a) ground-state energy per rung $e_0$, 
(b) velocity $v$, 
(c) scaling dimension $x$ of the SU(2)$_2$ WNZW primary field (see text for details). 
Data are compa\-ti\-ble with CFT behavior and extrapolations obtain $x=0.375$ and the central charge $c=1.52$.}
\label{fig:CFT_N2_J1}
\end{figure}

In principle, ED simulations can also give insight into the nature of this phase transition~\cite{Hijii2002}. 
Indeed, the central charge $c$ can be extracted from the finite-size corrections of the ground-state energy per rung $e_0(L)=E_0/L$ \cite{cardy,affleckc}: 
\begin{equation}
e_0(L) \simeq e_\infty - \frac{\pi v c}{6 L ^2}
\end{equation}
where $v$ is the velocity, which can be extrapolated from \footnote{We note that there are different finite-size effects depending either $L$ is multiple of 4 or not.}: 
\begin{equation}
v(L)=\frac{L}{2\pi} \left(E(q=\frac{2\pi}{L})-E(q=0)\right)
\end{equation}
Numerical data, computed at the critical point $K=0.26$, are shown in Fig.~\ref{fig:CFT_N2_J1}~(a,b) and confirm the $1/L^2$ corrections to the ground-state energy per site. We extracted the value of the central charge $c=1.52$, which is in excellent agreement with the expected value $c=3/2$ of the SU(2)$_2$ WZNW model. 

However, as detailed below in the DMRG part, while we are measuring the correlation exponent, logarithmic corrections, coming from marginal operators, seem to be present and spoil the accuracy of some of the computations.
In such case, it can be simpler to compute the scaling dimension $x$ of the lowest excitation of the expected SU(2)$_2$ WZNW model by removing explicitly these logarithmic corrections~\cite{Affleck1989,Totsuka1995,Hijii2009}
\begin{equation}
x(L) = \frac{L}{8\pi v(L)} [ 3\Delta E_0 (S=1) + \Delta E_0 (S=0)]
\end{equation}
where $\Delta E_0(S=0,1)$ are respectively the singlet and triplet gaps. Data are shown in Fig.~\ref{fig:CFT_N2_J1}~(c) and while finite-size effects are non-monotonic, they are relatively weak for large clusters and our numbers are compatible with the expected value $x=3/8$. 

\begin{figure}[ht]
\centering
\includegraphics[width=0.95\columnwidth,clip]{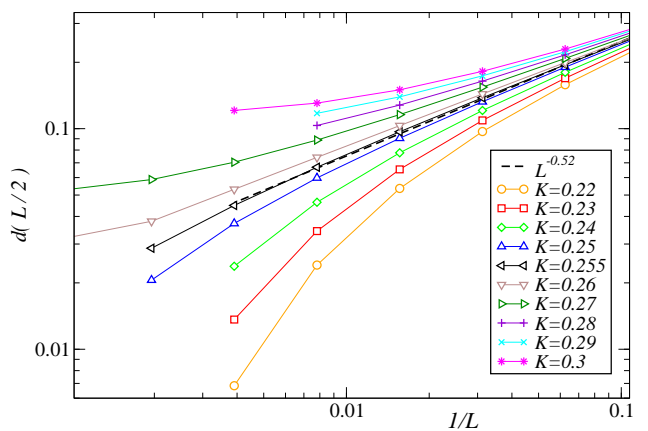}
\caption{(Color online) $N=2$ and $J_\perp=1$: dimerization at the center for $N=2$ for various values of $K$. The critical point corresponds to a power-law decay (see fit, dashed line) for $K=0.255$ (black).}
\label{fig:dimerOP_N2_J1}
\end{figure}

We now turn to DMRG simulations to confirm our findings. 
In order to detect the occurrence of a dimerization pattern, 
we measure the local dimerization at the center in the first chain, i.e. $d(L/2)=\langle {\bf S}_{1,L/2} \cdot {\bf S}_{1,L/2+1}\rangle$, and study its scaling versus length $L$. 
Indeed, according to the low-energy analysis of Sec.~\ref{sec:lowEnergy}, we expect this order parameter to decay as a power law at the transition, while it should converge exponentially to zero or to a constant below or above the transition.
In Fig.~\ref{fig:dimerOP_N2_J1}, we plot this order parameter versus system size on a log-log plot and it leads to an accurate determination of the critical point at $K/J_\perp=0.255$, which is in excellent agreement with our ED result. 

\begin{figure}[ht]
\centering
\includegraphics[width=0.95\columnwidth,clip]{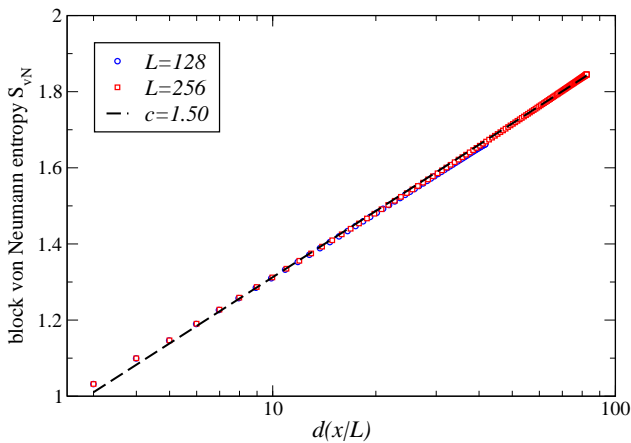}
\caption{(Color online) $N=2$ and $J_\perp=1$: von Neumann entropy versus conformal distance $d(x|L)$ at the critical point for $K=0.255$.} 
\label{fig:entropy_N2_J1}
\end{figure}

In order to characterize this critical point, we compute the entanglement entropy for such parameters. Indeed it is well known~\cite{Calabrese2004} that, for a critical point described by a CFT with central charge $c$, the von Neumann entropy behaves as
\begin{equation}
S_{vN}(\ell) = \frac{c}{6} \ln d(\ell|L)
\end{equation}
where $d(\ell|L)=(L/\pi) \sin (\pi \ell/L)$ is the conformal distance and OBC are used. 
Fig.~\ref{fig:entropy_N2_J1} shows that our data can be fitted with $c=1.50$, which is in perfect agreement with the expected value $c=3/2$ for an SU(2)$_2$ WZNW CFT. 

\subsubsection{$N=2$ ferromagnetic rung coupling case: $J_\perp=-1.0$}

\begin{figure}[ht]
\centering
\includegraphics[width=0.95\columnwidth,clip]{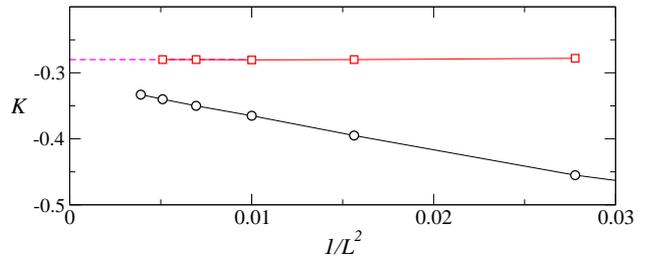}
\caption{(Color online) $N=2$ and $J_\perp=-1$: crossing point with PBC (circles) and TBC (squares) versus system size (see text for definitions). The most precise extrapolations are obtained using TBC data and lead to a critical point at $K=-0.28$ (dashed line).}
\label{fig:ED_N2_J1_ferro}
\end{figure}

In the case of a ferromagnetic rung coupling $J_\perp=-J_\parallel$, we need to consider $K<0$ in order to induce a transition from the Haldane phase to the UD phase. 
Again, this phase transition is detected using ED by comparing the first singlet and triplet excited states and plotting this crossing value versus system size, as shown in  Fig.~\ref{fig:ED_N2_J1_ferro}. However, since finite-size effects are quite strong with PBC, we prefer again to use TBC and compare eigenstates with quantum numbers $(+1,+1,-1)$ $(+1,-1,+1)$ (see definitions in Sec.~\ref{sec:N2J1}).  
Results for this level crossing are shown in Fig.~\ref{fig:ED_N2_J1_ferro} and TBC allow a more precise determination for the critical value at $K=-0.28$.

\begin{figure}[ht]
\centering
\includegraphics[width=0.95\columnwidth,clip]{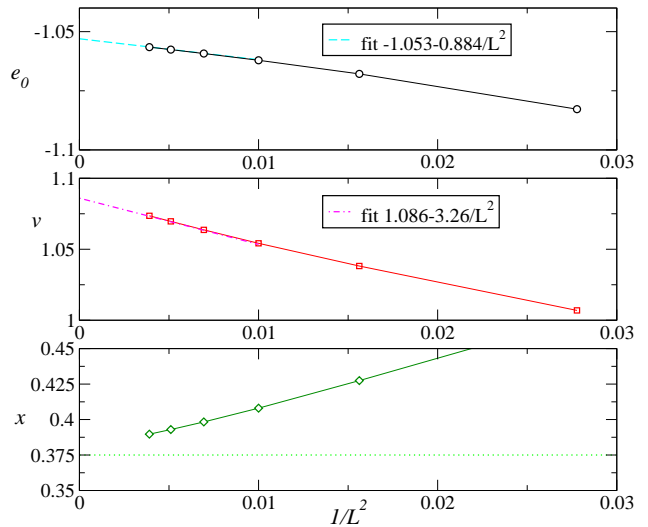}
\caption{(Color online) $N=2$ and $J_\perp=-1$: finite-size scaling of various quantities at the critical point $K=-0.28$: 
(a) ground-state energy per rung $e_0$, 
(b) velocity $v$, 
(c) scaling dimension $x$ of the SU(2)$_2$ WNZW primary field (see text for details). 
Data are compatible with CFT behavior and extrapolations obtain $x=0.376$ and the central charge $c=1.55$.}
\label{fig:CFT_N2_Jm1}
\end{figure}

Similarly to what we have done in the previous section for $J_\perp>0$, we fix this critical value and compute the ground-state energies, velocities and gaps in order to determine the CFT quantities. Data shown in Fig.~\ref{fig:CFT_N2_Jm1} allow to determine the central charge $c=1.55$ and the primary field dimension $x=0.376$ which are in excellent agreement with our prediction of a SU(2)$_2$ WZNW universality class. 

\begin{figure}[ht]
\centering
\includegraphics[width=0.95\columnwidth,clip]{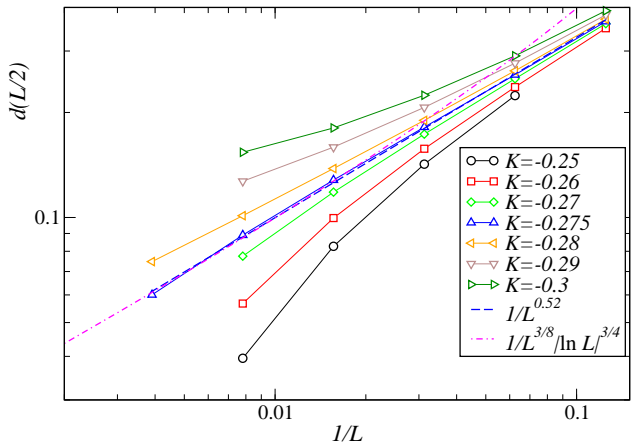}
\caption{(Color online) $N=2$ and $J_\perp=-1$: dimerization at the center for various $K$. Critical point is found at $K/J_\parallel=-0.275$ with a power-law behavior, possibly with logarithmic corrections (see text).}
\label{fig:dimerOP_N2_Jm1}
\end{figure}

We now turn to DMRG simulations in order to confirm our findings on larger scales. 
By computing the dimerization at the center (see Fig.~\ref{fig:dimerOP_N2_Jm1}), we can locate the transition for $K/J_\parallel \simeq -0.275$, which is in excellent agreement with our ED estimate. 
At the critical point the dimerization decays as a power-law $L^{-\alpha}$ with $\alpha= 0.52$, which does not agree with the expected $3/8$ exponent (see Eq.~(\ref{correldimer})). 
We believe that this disagreement is due to the existence of logarithmic corrections since a reasonable fit using Eq.~(\ref{correldimer}) can be achieved with $x=3/8$, as also found using ED where we had removed these logarithmic corrections.

\begin{figure}[ht]
\centering
\includegraphics[width=0.95\columnwidth,clip]{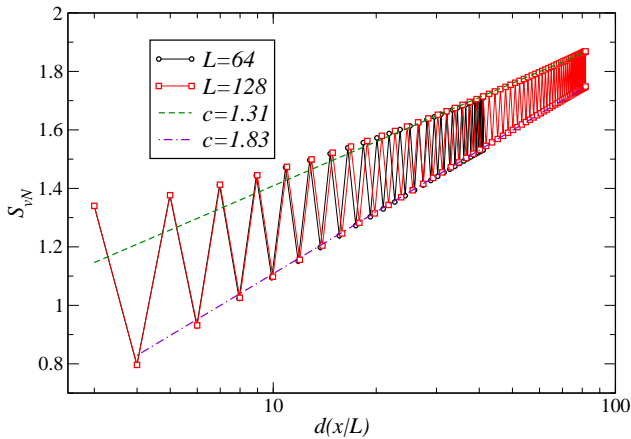}
\caption{(Color online) $N=2$ and $J_\perp=-1$: von Neumann entropy versus conformal distance $d(x|L)$ at the critical point $K=-0.275$. Oscillations prevent from a reliable fit of the central charge.} 
\label{fig:entropy_N2_Jm1}
\end{figure}

\begin{figure}[ht]
\centering
\includegraphics[width=0.95\columnwidth,clip]{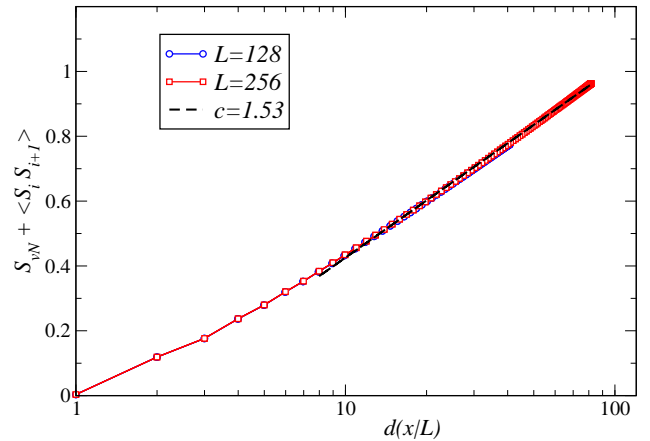}
\caption{(Color online) $N=2$ and $J_\perp=-1$: same data as in Fig.~\ref{fig:entropy_N2_Jm1} after removing the oscillations due to the bond modulations using $S_{vN}+ \langle S_i \cdot S_{i+1}\rangle$, see Eq.~(\ref{eq:newfit}). The fit of the central charge (dashed line) then gives $c=1.53$.}
\label{fig:entropy2_N2_Jm1}
\end{figure}

Next we would like to characterize further this critical point using the scaling of the von Neumann entropy. However, while the use of OBC improves convergence for DMRG, it can give rise to so-called Friedel oscillations (see Fig.~\ref{fig:entropy_N2_Jm1}) which complicate the analysis. 
Nevertheless, these oscillations are known to originate from the bond modulations~\cite{Fagotti2011,Laflorencie2006}, so we propose to use the following fitting form~\cite{Lavarelo2011}:
\begin{equation}
S_{vN}(i) = A+ \frac{c}{6}\ln d(i|L) + B \langle {\bf S}_i \cdot {\bf S}_{i+1} \rangle
\label{eq:newfit}
\end{equation}
where $\langle {\bf S}_i \cdot {\bf S}_{i+1} \rangle = \sum_a \langle
{\bf S}_{a,i} \cdot {\bf S}_{a,i+1} \rangle$ is the sum of the bond
contributions.  At the critical point, we have found $B=-1$ as the best
parameter, as it completely removes the oscillations (see
Fig.~\ref{fig:entropy2_N2_Jm1}), allowing for a precise determination
of $c=1.53$ in excellent agreement with the $c=3/2$ analytic
prediction.  Note that we prefer this fitting procedure instead of
using PBC which also remove oscillations but
require much larger numerical effort. Indeed the number of kept states
should be much larger for a similar accuracy, typically $m^2$ where $m$
is the number of kept states with OBC. Let us also mention that 
other proposals have been made to avoid fitting the oscillations~\cite{Xavier2010}, but they only use part of the available data.

\subsection{Three-leg ladders ($N=3$)}
We now turn to the three-leg geometry ($N=3$) using a similar analysis. Note that since chains are not equivalent, we have checked that the dimerization measured on chain 1 gives similar results with the one averaged over all chains.
Let us also remind that since $N$ is odd, the system is critical in the absence of cyclic exchange $K=0$ which implies that the transition is now from a gapless phase to a gapped one. 

\subsubsection{$N=3$ antiferromagnetic rung coupling case: $J_\perp=1.0$}
In this case we cannot use TBC in ED simulation to get an accurate estimate of the critical point. But even with PBC, we do not observe any level crossing between the lowest singlet and triplet with momentum $\pi$ (with respect to the ground-state) for lengths $L$ up to 12. Therefore we have to determine first the critical point using DMRG before performing an ED analysis. 

\begin{figure}[ht]
\centering
\includegraphics[width=0.95\columnwidth,clip]{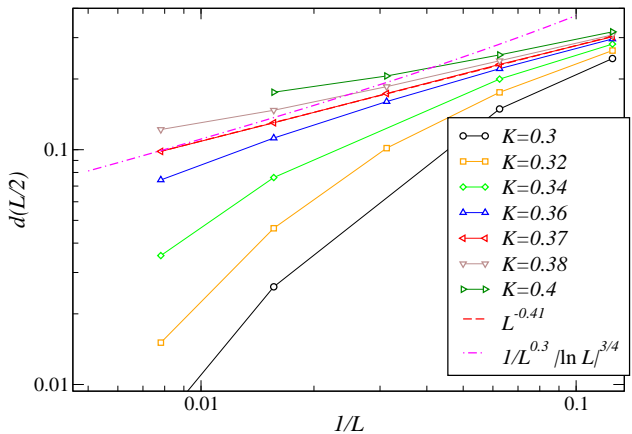}
\caption{(Color online) $N=3$ and $J_\perp=1$: dimerization at the center for various $K$. Power-law decay allows to locate the critical point for $K/J_\parallel=0.37$.}
\label{fig:dimerOP_N3_J1}
\end{figure}

As done before, we measure the dimerization in the middle of the first chain. Data are shown in Fig.~\ref{fig:dimerOP_N3_J1} and do indicate a change of concavity for $K/J_\parallel = 0.37$. Note that this is not entirely expected since the dimerization should obey a power-law in the whole critical phase for $K<K_c$ and converge exponentially to a finite value in the gapped SD phase for $K>K_c$ only. Nevertheless, if we take this heuristic criterion to detect the quantum phase transition, we get consistent results with other approaches (see for instance below for the ferromagnetic case where DMRG and ED agree on the location of the critical point). Let us also mention that this change of concavity at the transition was also observed recently for a similar phase transition in spin$-3/2$ chain with three-site interaction.~\cite{mila}

At the critical point, we can either fit our data with a power-law $1/L^{0.41}$, but the expected logarithmic corrections (see Eq.~(\ref{correldimer})) could also give a reasonable fit $1/(L^{0.3} |\ln L|^{3/4})$ (see Fig.~\ref{fig:dimerOP_N3_J1}). 

\begin{figure}[ht]
\centering
\includegraphics[width=0.95\columnwidth,clip]{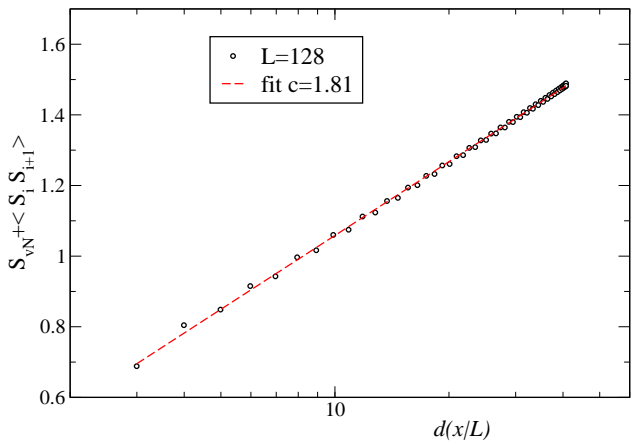}
\caption{(Color online) $N=3$ and $J_\perp=1$: von Neumann entropy versus conformal distance $d(x|L)$ at the critical point for various sizes. Again, oscillations were removed (see text).} 
\label{fig:entropy_N3_J1}
\end{figure}

Now that the critical point is located, we can study the scaling of the von Neumann entropy, and as before, we remove oscillations by adding the bond energies and plotting $S_{vN}(i) + \langle {\bf S}_i \cdot {\bf S}_{i+1} \rangle$ versus the conformal length $d(x|L)$ (see Fig.~\ref{fig:entropy_N3_J1}). A simple fit gives a central charge $c=1.81$ which is in excellent agreement with the CFT prediction $c=9/5=1.8$. 

\begin{figure}[ht]
\centering
\includegraphics[width=0.95\columnwidth,clip]{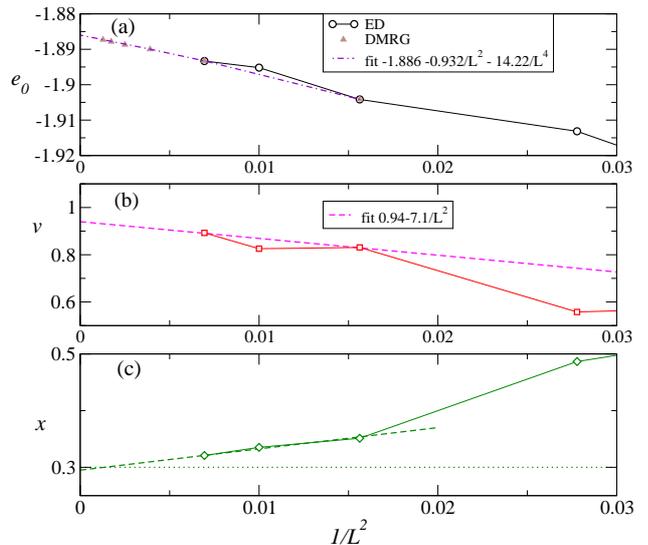}
\caption{(Color online) $N=3$ and $J_\perp=1$: finite-size scaling of various quantities at the critical point $K=0.37$: (a)
  ground-state energy per rung $e_0$, (b) velocity $v$, (c) scaling dimension $x$ of the primary
  CFT field (see text for details). }
\label{fig:CFT_N3_J1}
\end{figure}

Assuming now the critical point to be at $K=0.37$, we can perform a similar ED analysis as for $N=2$. Data are shown in Fig.~\ref{fig:CFT_N3_J1}. Unfortunately, strong finite-size effects, as well as different scalings required for $L=4p$ and $L=4p+2$ ladders, do not allow for accurate extrapolations of $e_0$ and $c$. Nevertheless, if we also perform DMRG with PBC on moderate $L$ (in which case we have kept up to $m=4000$ states for $L$ up to 28), we can obtain a reasonable fit of $e_0$ versus $1/L^2$ (see Fig.~\ref{fig:CFT_N3_J1}~(a)). From this, we extracted $c=1.89$, which is in reasonable agreement with our previous finding and expectation. 
Note that, since we have used variational DMRG energies, one could suspect that either increasing the number of states kept or extrapolating with the discarded weight, energies will be slightly lower for the largest sizes, thus leading to a reduced slope and $c$ value. 

However, ED also allows to extract the scaling dimension $x$ by getting rid of the logarithmic corrections. Our numerical data support the expected $x=3/10$ with an excellent accuracy (see Fig.~\ref{fig:CFT_N3_J1}~(c)).

\subsubsection{$N=3$ ferromagnetic rung coupling case: $J_\perp=-1.0$}

\begin{figure}[ht]
\centering
\includegraphics[width=0.95\columnwidth,clip]{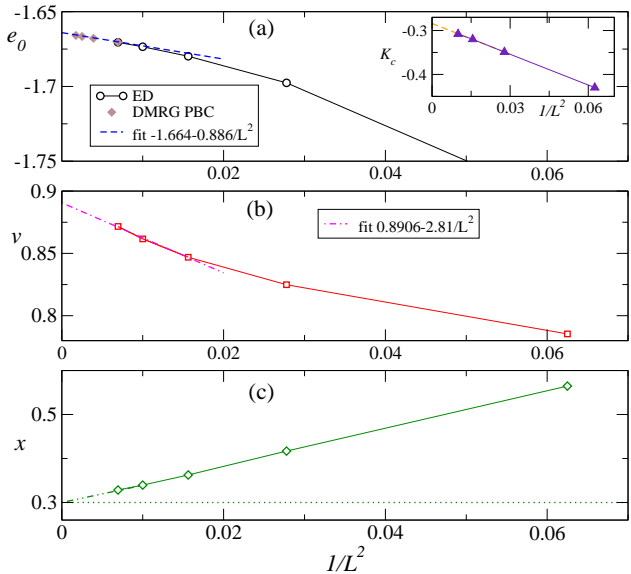}
\caption{(Color online) $N=3$ and $J_\perp=-1$: finite-size scaling of various quantities at the critical point $K=-0.28$: (a)
  ground-state energy per rung $e_0$, (b) velocity $v$, (c) scaling dimension $x$ of the primary
  CFT field (see text for details). Inset: scaling of the level crossing versus system size.}
\label{fig:CFT_N3_Jm1}
\end{figure}

When considering ferromagnetic rung coupling, ED simulations show a level crossing between the lowest singlet and triplet excitations at momentum $\pi$ with respect to the ground-state momentum. This enables a determination of the critical point at $K_c=-0.284$ (see the inset of Fig.~\ref{fig:CFT_N3_Jm1}), which is exactly the same as previously found for $N=2$. By performing a finite-size scaling analysis at fixed $K=-0.28$, one can extract CFT quantities as explained above. Fig.~\ref{fig:CFT_N3_Jm1}~(a-c) present finite-size scaling behavior of $e_0$, $v$ and $x$. As in the previous case, size limitation does not allow an accurate estimate of the slope of the energy, but still we note that the extrapolation of $x(L)$ is rather smooth and leads to $x=0.3$ as predicted. In order to get an estimate of the central charge, we have also performed DMRG with PBC (up to $L=24$) in order to compute $e_0$. Fitting $e_0$ and $v$ allows to extract an estimate $c=1.90$ which is not far from the expected $9/5$. 

\begin{figure}[ht]
\centering
\includegraphics[width=0.95\columnwidth,clip]{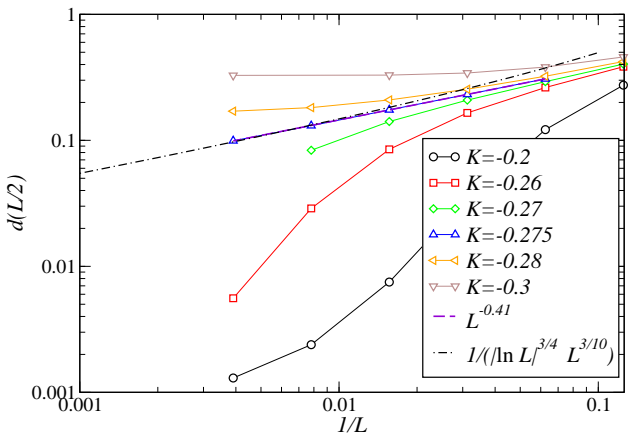}
\caption{(Color online) $N=3$ and $J_\perp=-1$: dimerization at the center for various $K$. The critical point corresponds to a power-law decay for $K/J_\parallel=-0.275$.}
\label{fig:dimerOP_N3_Jm1}
\end{figure}

Turning now to DMRG with OBC, the quantum phase transition from the critical phase to the UD phase can be located as usual on Fig.~\ref{fig:dimerOP_N3_Jm1}. From the power-law decay of the dimerization, we determined the transition at $K/J_\parallel = -0.275$, a value that is extremely close from our ED estimate, thus giving us confidence in its accuracy. Indeed, we remind that since dimerization should obey a power-law in \emph{the whole} critical phase, it is not obvious that there is a change of concavity at the transition. Nevertheless, if we try to fit the dimerization decay at the transition, we get exactly the same behavior as in the antiferromagnetic case, namely that we can fit either with 
a power-law $L^{-0.41}$ (exact same exponent as for $J_\perp,\, K >0$) or with the expected behavior from Eq.~(\ref{correldimer}) $1/(L^{0.3} |\ln L|^{3/4})$.

\begin{figure}[ht]
\centering
\includegraphics[width=0.95\columnwidth,clip]{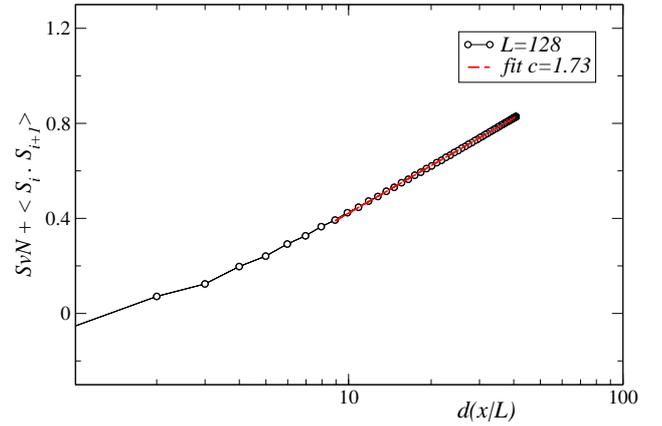}
\caption{(Color online) $N=3$ and $J_\perp=-1$: von Neumann entropy plus bond energy versus conformal distance $d(x|L)$ at the critical point $K=-0.275$ for $L=128$.}
\label{fig:entropy_N3_Jm1}
\end{figure}

Since the von Neumann entropy at the transition exhibits large oscillations (see details in Sec.~\ref{sec:N2J1}), we use the same technique as before and add the total bond energies. In Fig.~\ref{fig:entropy_N3_Jm1}, we do observe that oscillations have disappeared and a rather good fit can be performed. We obtain $c=1.73$, in good agreement with the expected $c=9/5=1.8$. 

\subsection{$N=4$}
\subsubsection{$N=4$ antiferromagnetic rung coupling case: $J_\perp=1.0$}

\begin{figure}[ht]
\centering
\includegraphics[width=0.95\columnwidth,clip]{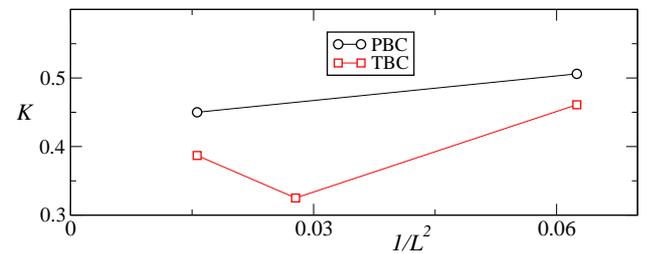}
\caption{(Color online) $N=4$ and $J_\perp=1$: Crossing point with PBC and TBC versus system size (see text for definitions). 
Extrapolations are difficult but point towards a critical value between $0.3< K_c <0.5$. 
}
\label{fig:ED_N4_J1_antiferro}
\end{figure}

We start by performing ED simulations and using the crossing point either with PBC or TBC similarly to the $N=2$ case (see Sec.~\ref{sec:N2J1} for details). Results are shown in Fig.~\ref{fig:ED_N4_J1_antiferro} and do not allow for a precise determination of $K_c$ which we estimate to be between 0.3 and 0.5. Therefore we cannot perform a CFT analysis at this stage. 

\begin{figure}[ht]
\centering
\includegraphics[width=0.95\columnwidth,clip]{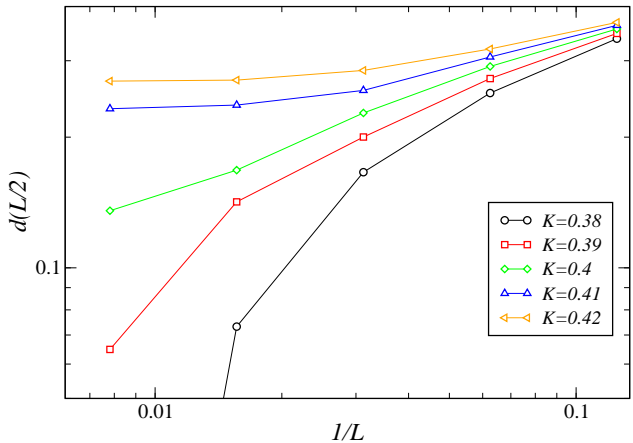}
\caption{(Color online) $N=4$ and $J_\perp=1$: dimerization at the center for various $K$. Critical point corresponds to a power-law decay for $K/J_\parallel \simeq 0.4$}
\label{fig:dimerOP_N4_J1}
\end{figure}

By performing DMRG simulations and plotting the dimerization at the center (see Fig.~\ref{fig:dimerOP_N4_J1}), we can locate the transition precisely for $K/J_\parallel \simeq 0.4$. Note this case is similar to $N=2$ in the sense that both phases are gapped and a change of concavity is expected at the transition. 

At the critical point the dimerization decays as a power-law $L^{-\alpha}$ with $\alpha \simeq 0.5$, which does not agree with the expected $3/8$ exponent. We believe that this disagreement is due to the existence of logarithmic corrections as had been observed for other values of $N$.

\begin{figure}[ht]
\centering
\includegraphics[width=0.95\columnwidth,clip]{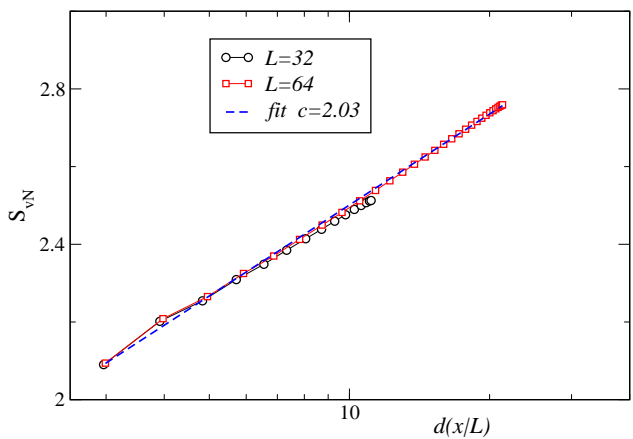}
\caption{(Color online) $N=4$ and $J_\perp=1$: von Neumann entropy versus conformal distance $d(x|L)$ at the critical point $K=0.4$ for various sizes.} 
\label{fig:entropy_N4_J1}
\end{figure}

At this critical point, by keeping a large number of states ($m=4800$ for $L=64$), we are able to get a very nice entropy plot (see Fig.~\ref{fig:entropy_N4_J1}) and extract $c=2.0$ in perfect agreement with low-energy prediction. 

\begin{figure}[ht]
\centering
\includegraphics[width=0.95\columnwidth,clip]{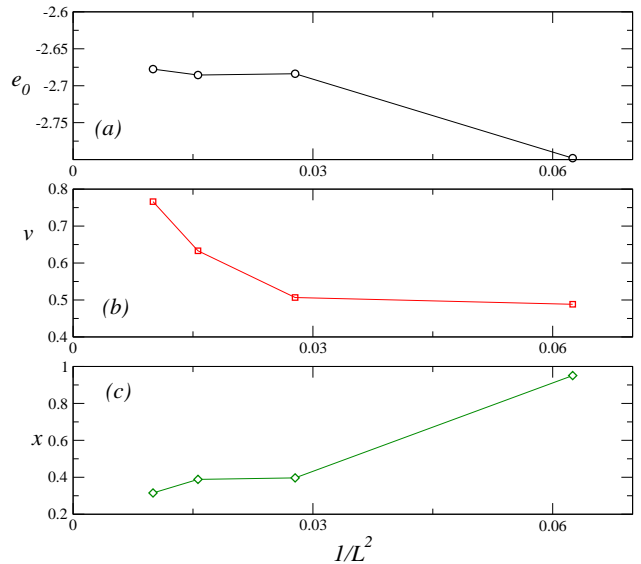}
\caption{(Color online) $N=4$ and $J_\perp=1$: finite-size scaling of various quantities at the critical point $K=0.4$: (a)
  ground-state energy per rung $e_0$, (b) velocity $v$, (c) scaling dimension $x$ of the primary
  CFT field (see text for details). }
\label{fig:CFT_N4_J1}
\end{figure}

Going back now to ED simulations  at $K=0.4$ (up to $4\times 10$ ladder), we can perform a similar analysis as before. Data are shown in Fig.~\ref{fig:CFT_N4_J1}. Unfortunately, strong finite-size effects inhibit us for accurate extrapolations of $e_0$, $v$ or $c$. Similarly, the scaling dimension $x$ does not show  a sufficiently smooth behavior to extract an accurate estimate, but it is reasonable to extrapolate it between 0.2 and 0.3, which could be compatible with the expected $1/4$ value. Clearly, we are severely limited in system size and it would be useful to be able to remove logarithmic corrections in DMRG simulations. 

\subsubsection{$N=4$ ferromagnetic rung coupling case: $J_\perp=-1.0$}

\begin{figure}[ht]
\centering
\includegraphics[width=0.95\columnwidth,clip]{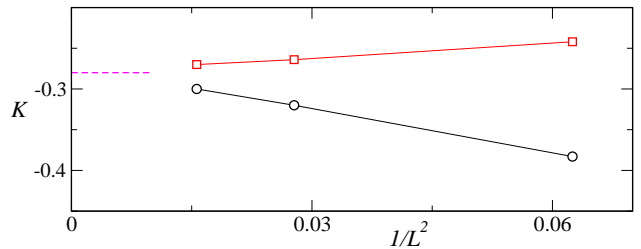}
\caption{(Color online) $N=4$ and $J_\perp=-1$: crossing point with PBC and TBC versus system size (see text for definitions). 
Both PBC and TBC allow for an accurate extrapolation to $K_c=-0.28$. 
}
\label{fig:ED_N4_J1_ferro}
\end{figure}

\begin{figure}[ht]
\centering
\includegraphics[width=0.95\columnwidth,clip]{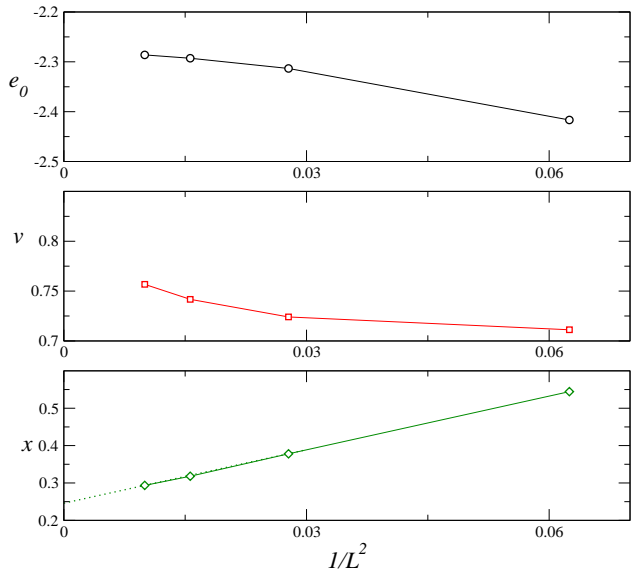}
\caption{(Color online) $N=4$ and $J_\perp=-1$: finite-size scaling of various quantities at the critical point $K=-0.28$: 
(a) ground-state energy per rung $e_0$, 
(b) velocity $v$, 
(c) scaling dimension $x$ of the primary CFT field.}
\label{fig:CFT_N4_Jm1}
\end{figure}

ED crossing results are shown in Fig.~\ref{fig:ED_N4_J1_ferro} and the use of either PBC or TBC data allows to determine the critical point $K=-0.28$, again identical to what we had found for $N=2$ and $N=3$. By fixing this value and analyzing our spectra (using up to $4\times 10$ ladders), we can perform a CFT analysis similarly to what has been done for the other $N$. Data are shown in Fig.~\ref{fig:CFT_N4_Jm1}. As for $N=3$, size limitation does not allow for a determination of $c$, however the behavior of $x(L)$ is smoother (especially in contrast to the case $J_\perp=1$) and leads to an estimate $x=0.246$ in excellent agreement with the expected value $1/4$ for $k=4$ SU(2) WZNW. 

\begin{figure}[ht]
\centering
\includegraphics[width=0.95\columnwidth,clip]{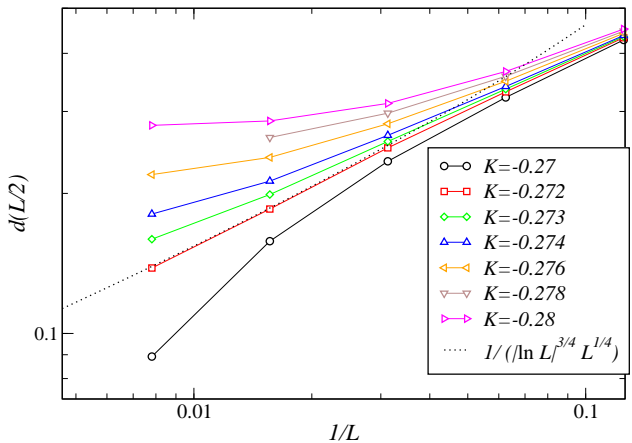}
\caption{(Color online) $N=4$ and $J_\perp=-1$: dimerization at the center for various $K$. Critical point corresponds to $K/J_\parallel=-0.272$ with a power-law decay with exponent $x=1/4$ up to logarithmic corrections.}
\label{fig:dimerOP_N4_Jm1}
\end{figure}

As usual, we can also compute the dimerization at the center by DMRG (see Fig.~\ref{fig:dimerOP_N4_Jm1}), which allows us 
to locate the transition precisely at $K=-0.272$, in excellent agreement with our ED finding. In particular, it confirms that this critical value seems independent from $N$. Note that, at the transition, a simple power-law fit would lead to $L^{-0.41}$ which is not in agreement with the expected $x=0.25$; however, taking into account logarithmic corrections, we can perform a quite satisfactory $L^{-0.25}/|\ln{L}|^{3/4}$ fit. 

\begin{figure}[ht]
\centering
\includegraphics[width=0.95\columnwidth,clip]{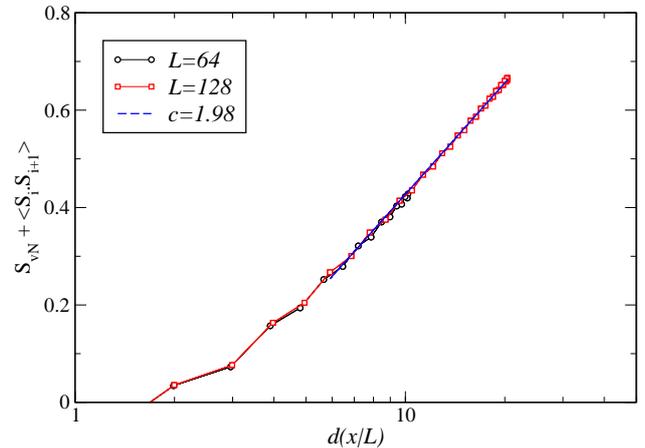}
\caption{(Color online) $N=4$ and $J_\perp=-1$: von Neumann entropy plus bond energy versus conformal distance $d(x|L)$ at the critical point $K=-0.272$ for various sizes.}
\label{fig:entropy_N4_Jm1}
\end{figure}

Playing again with fitting the von Neumann entropy by removing oscillations (simply by adding the total bond energies, see Fig.~\ref{fig:entropy_N4_Jm1}) allows to get an excellent value for $c$ in agreement with $c=2.0$. 


\section{Concluding remarks}
\label{sec:Concluding}

In summary, we have investigated the nature of the quantum phase
transitions of $N$-leg spin ladders with a cyclic four-spin exchange
interaction by means of complementary techniques.  SD and UD dimerized
phases emerge from this ring exchange on top of the conventional
phases of the $N$-leg spin ladders. Several quantum phase transitions
have been studied here depending on the sign of the ring-exchange
interaction $K$ and the parity of $N$.  When $N$ is even, we can
investigate the transition between the RS (respectively Haldane) phase
and the SD (respectively UD) phase when $K>0$ (respectively $K <
0$). In contrast, when $N$ is odd, a transition occurs between the
$c=1$ Heisenberg gapless phase and the dimerized phases (SD and UD
depending on the sign of $K$).

Using a low-energy approach, when the chains are weakly coupled, we
find that all these quantum phase transitions are described by an
SU(2)$_N$ CFT with central charge $c=3N/(N+2)$. This result
generalizes the findings of Ref. \onlinecite{shura} for $N=2$ where a
SU(2)$_2$ emerging quantum criticality has been revealed by means of a
Majorana-fermion approach, which is very specific to the $N=2$
case. In this respect, we have shown that the quantum phase
transitions, driven by four-spin exchange interactions, belong to the
SU(2)$_N$ universality class when $N=2,3,4$.  For $N>4$, a relevant
perturbation is likely to be generated and to drive the system away
from criticality. A first-order phase transition is then expected for
$N>4$ although further work is clearly called for to confirm or infirm
this result.

In close parallel to this field-theory analysis, we have investigated
numerically the phase transitions for $N=2,3,4$ by means of ED and
DMRG approaches. As we have seen, these techniques give a precise
location of the quantum phase transitions for all signs of $K$. The
position of the transition does not vary too much as a function of $N$
when $J_{\perp} > 0$, and seems to be independent of $N$ when
$J_{\perp} <0$. Surprisingly enough, the estimate of the transition
obtained within the low-energy approach is in rather good agreement
with the numerical results.  The nature of the quantum phase
transitions was then obtained by extracting numerically the central
charge $c$ and the scaling dimension $x$ of the lowest primary field of the
SU(2)$_N$ CFT.   Let us emphasize that while DMRG is the tool of choice to measure the central charge $c$, the presence of logarithmic correction prevents a reliable determination of $x$; on the contrary, $x$ can  be extracted accurately from ED data as shown in our study. 
In all cases, we found a very good agreement with the
prediction of the low-energy approach according to which the quantum
phase transitions should belong to the SU(2)$_N$ universality class
for $N \le 4$. 
In this respect, the $N=3$ case is intriguing due to
the prediction of a transition between the standard gapless $c=1$
phase and the dimerized phases (SD and UD phases) that is not of the
BKT type with central charge $c=1$, as for the $J_1-J_2$ Heisenberg
spin$-1/2$ chain, but rather an SU(2)$_3$ transition with the
emergence of non-trivial critical modes with fractional central charge
$c=9/5$. All our results are summarized in Table~\ref{table}.

\renewcommand{\arraystretch}{1.2}
\begin{table}[h!]
\begin{tabular}{c c| c c c}
    \hline\hline
    \multicolumn{2}{c}{} & $N=2$ & $N=3$ & $N=4$ \\
    \hline
    \multirow{3}{*}{low-energy analysis} & $K/J_\perp$ & $\pi^2/36\simeq 0.274$ &  $\pi^2/36$ &  $\pi^2/36$ \\
    & $c$ & 3/2 & 9/5 & 2 \\
    & $x$ & 3/8 & 3/10 & 1/4 \\
    \hline
    numerical & $K/J_\perp$ & 0.255 &  0.37 &  0.4 \\
    \multirow{2}{*}{$J_\perp=J_\parallel$} & $c$ & 1.52 & 1.81 & 2.0 \\
    & $x$ & 0.375 & 0.3 & 0.2-0.3 \\
    \hline
    numerical & $K/J_\perp$ & 0.275 &  0.28 &  0.27-0.28 \\
    \multirow{2}{*}{$J_\perp=-J_\parallel$} & $c$ & 1.55 & 1.73 & 1.98 \\
    & $x$ & 0.376 & 0.3 & 0.246 \\
    \hline\hline
\end{tabular}
\caption{\label{table}
Summary of our results showing the existence of a quantum phase transition in $N$-leg spin-1/2 ladder with ring exchange $K$. Low-energy analysis predicts a transition for $K/J_\perp=\pi^2/36 \simeq 0.274$ in the SU(2)$_N$ universality class for $N=2,3,4$ with central charge $c=3N/(N+2)$ and scaling dimension $x=3/(2N+4)$. Numerical analysis performed by ED and DMRG simulations (see text) provides a very good confirmation of the nature of this quantum phase transition. 
}
\end{table}

The field-theory and numerical approaches presented in this paper
therefore show that four-spin exchange interaction may generate an
exotic emerging quantum criticality in one dimension.  In this
respect, there is the question whether our analytical approach can
shed some light on the deconfined quantum criticality universality
class between Néel and UD phases in two dimensions.~\cite{senthil_1}
Unfortunately, we are not able to determine the nature of the phase
transition between the RS (or the $c=1$ gapless phase) and UD phases
for $K<0$ when $N>2$ within our low-energy approach. A
phenomenological approach, describing the 2D system as an infinite
array of 1D spin$-1/2$ Heisenberg chains, predicts that the N\'eel-UD
transition is governed by an anisotropic O(4) non-linear $\sigma$ model
with a topological term.~\cite{senthilfisher06} Our work is more
relevant to the investigation of the transition between the Néel phase
and a staggered VBS which has been observed numerically in the
spin$-1/2$ Heisenberg model with a ring-exchange on the square
lattice.~\cite{Lauchli05} An estimation of the location of the
transition by means of ED calculations is $K/J \sim
0.4$ \footnote{A.~L{\"a}uchli and P.~Sindzingre, private
  communications.}  which is close to our numerical findings in the
ladder limit. The nature of the universality class of the transition
is unknown but it seems to be different from the deconfined
criticality paradigm.~\cite{balentsxu} Furthermore a first-order
transition has been found in the Quantum Monte Carlo simulations of
the spin$-1/2$ Heisenberg model with a special six-spin exchange
interaction .~\cite{sandvik1order} Our field-theory approach, combined
with the ideas presented in Ref.~\onlinecite{senthilfisher06}, seems
to suggest that the transition in two dimensions can be described in
terms of an anisotropic O(4) non-linear $\sigma$ model.  In the future,
it will be very meaningful to pursue this approach starting from
coupled spin chains to investigate the nature of two dimensions
Néel-SD transition, in particular in the easy-plane limit to make
connections with the phenomenological theory of
Ref.~\onlinecite{balentsxu}.

\section*{Acknowledgements}
The authors would like to thank  F. Alet, F. Mila, H. Nonne, E. Orignac, G. Roux,  and K. Totsuka  for insightful discussions. 
Numerical simulations were performed at CALMIP and GENCI.


\appendix

\section{Continuum limit}

In this Appendix, we provide the technical details for 
the derivation of the continuum limit of model (\ref{nlegringexchange}).
The ring-exchange operator $P_{n+1,n}$,  which
performs a cyclic permutation of the spins ${\bf S}_{a, n},{\bf
S}_{a,n+1},{\bf S}_{a+1,n+1},{\bf S}_{a+1,n}$ between two consecutive chains,
can be written in terms of quadratic and biquadratic products
of spins as \cite{Brehmer99,sakai}
\begin{eqnarray}
  \label{eq:ring-operator}
&&  P_{n,n+1} + P_{n,n+1}^{-1} =\frac 1 {4} + {\bf S}_{a,n}\cdot {\bf S}_{a+1,n} +{\bf
    S}_{a,n+1}\cdot {\bf S}_{a+1,n+1} 
    \nonumber \\
    &+& {\bf S}_{a,n}\cdot{\bf
    S}_{a,n+1}+ {\bf S}_{a+1,n}\cdot{\bf S}_{a+1,n+1} +{\bf S}_{a,n}\cdot
  {\bf S}_{a+1,n+1} \nonumber \\
  &+& {\bf S}_{a+1,n}\cdot{\bf S}_{a,n+1} + 4 ({\bf
    S}_{a,n}\cdot {\bf S}_{a+1,n})({\bf S}_{a,n+1}\cdot {\bf S}_{a+1,n+1})
     \nonumber \\
  &+& 4({\bf S}_{a,n}\cdot{\bf
    S}_{a,n+1})({\bf S}_{a+1,n}\cdot{\bf S}_{a+1,n+1}) 
  \nonumber \\  
    &-& 4 ({\bf S}_{a,n}\cdot
  {\bf S}_{a+1,n+1})({\bf S}_{a+1,n}\cdot{\bf S}_{a,n+1}).
\end{eqnarray}
The spin ladder with a ring-exchange (\ref{nlegringexchange})
is thus a  particular case of a general Hamiltonian 
corresponding to a ladder with quadratic and biquadratic spin-spin
interactions:
\begin{eqnarray}
&& {\cal H}_{\rm gen} = \left(J_{\parallel} + J_\ell\right) 
\sum_{n}  \sum_{a=1}^{N} {\bf S}_{a, n}
\cdot {\bf S}_{a,n+1} 
\nonumber \\
&+& J_{r} \sum_n  \sum_{a=1}^{N-1}
{\bf S}_{a, n}\cdot{\bf S}_{a+1,n}  
\nonumber \\
&+& J_d \sum_n  \sum_{a=1}^{N-1} \left(
{\bf S}_{a, n} \cdot {\bf S}_{a+1,n+1}
+ {\bf S}_{a,n+1} \cdot {\bf S}_{a+1,n} \right)
\nonumber \\
&+& J_{rr} \sum_n \sum_{a=1}^{N-1} \left(
{\bf S}_{a,n} \cdot {\bf S}_{a+1,n}\right)
\left({\bf S}_{a,n+1} \cdot {\bf S}_{a+1,n+1}\right)
\nonumber \\
&+&  J_{\ell\ell} \sum_n  \sum_{a=1}^{N-1} \left(
{\bf S}_{a,n} \cdot {\bf S}_{a,n+1}\right)
\left({\bf S}_{a+1,n} \cdot {\bf S}_{a+1,n+1}\right) \nonumber \\ 
&+& J_{dd} \sum_n  \sum_{a=1}^{N-1} \left( {\bf S}_{a,n} \cdot {\bf S}_{a+1,n+1}
\right) \left( {\bf S}_{a,n+1} \cdot {\bf S}_{a+1,n}\right),
\label{laddergen} 
\end{eqnarray} 
with $J_{rr} = J_{\ell\ell} = - J_{dd} = 4K$, 
and $J_r = J_{\perp} + 2K, J_\ell = K, J_d= K$ for the spin-ladder with a ring-exchange (\ref{eq:ring-operator}).

In the following, we are going to investigate the continuum
limit of the model (\ref{laddergen}) in 
the weak coupling limit when $|J_\ell, J_r, J_d, J_{rr}, J_{\ell\ell} ,
J_{dd}| \ll J_{\parallel}$, i.e. in the vicinity of the decoupling point.
The continuum limit was done in the two-leg case, i.e. $N=2$, 
in Ref.~\onlinecite{gritsev}, but we believe that the identification of the coupling
constants of the effective Hamiltonian are not correct.
To perform such calculations, one needs a free-field representation of the 
fields of Eq.~(\ref{spinop})  in terms
a bosonic field $\Phi_a$ with chiral components 
$\Phi_{aR,L}$ \cite{bookboso}:
\begin{eqnarray}
{\bf n}_a &=& \frac{\lambda}{\pi a_0} \Big(\cos\big(\sqrt{2\pi}\;\Theta_a \big), \sin\big(\sqrt{2\pi}\;\Theta_a \big), 
- \sin\big(\sqrt{2\pi}\;\Phi_a \big)\Big)  
\nonumber \\
\epsilon_a &=& \frac{\lambda}{\pi a_0} 
\cos\left(\sqrt{2\pi}\; \Phi_a \right)
\nonumber \\
J_{a R,L}^z &=& \frac{1}{\sqrt{2\pi}} \; \partial_x \Phi_{a R,L} 
\nonumber \\
J_{a L}^{+} &=& \frac{1}{2\pi a_0} 
\exp\left(i \sqrt{8 \pi}\; \Phi_{a L} \right), 
\nonumber \\
J_{a R}^{+} &=& \frac{1}{2\pi a_0} 
\exp\left(-i \sqrt{8 \pi}\; \Phi_{a R} \right) ,
\label{bose}
\end{eqnarray}
where the bosonic fields are normalized according to:
\begin{eqnarray}
\langle \Phi_{a L} \left(z \right) 
\Phi_{b L} \left(0 \right) \rangle = - \frac{\delta^{ab}}{4 \pi} 
\ln  z \nonumber \\
\langle \Phi_{a R} \left(\bar z \right) 
\Phi_{b R} \left(0 \right) \rangle = - \frac{\delta^{ab}}{4 \pi} 
\ln  \bar z ,
\label{normali}
\end{eqnarray}
where $z= v\tau + i x$ ($\tau$ being the imaginary time) 
and the boson fields satisfy the following commutation
relations: $[\Phi_{a R} (x), \Phi_{b L} (y)] = i \delta^{ab}/4$.
Within this Abelian bosonization procedure, the WZNW field $g_a$
of Eqs. (\ref{staggop}, \ref{dimerizationop}) expresses as:
\begin{eqnarray}
g_a  = \frac{1}{\sqrt{2}}
\left(
\begin{array}{lccr}
 e^{-i \sqrt{2\pi} {\Phi}_a}  &
i  e^{-i \sqrt{2\pi} {\Theta}_a}  \\
i  e^{i \sqrt{2\pi} {\Theta}_a}   &
 e^{i \sqrt{2\pi} {\Phi}_a} 
\end{array} \right) .
\label{gbos}
\end{eqnarray}

This free-boson representation of the SU(2)$_1$ CFT 
is also useful to derive the different OPEs which will  
play a crucial role in the continuum limit of the 
general Hamiltonian (\ref{laddergen}):
\begin{eqnarray}
{J}_{L}^{\alpha} \left(z\right) {J}_{L}^{\beta} \left(w\right)
&\sim& \frac{\delta^{\alpha \beta}}{8 \pi^2 \left(z - w \right)^2}
+ i \epsilon^{\alpha \beta \gamma} \frac{{J}_{L}^{\gamma} \left(w\right)}{2 \pi 
\left(z - w\right)}
\nonumber \\
{J}_{R}^{\alpha} \left(\bar z\right) {J}_{R}^{\beta} \left(\bar w\right)
&\sim& \frac{\delta^{\alpha \beta}}{8 \pi^2 \left(\bar z - \bar w \right)^2}
+ i \epsilon^{\alpha  \beta \gamma} \frac{{J}_{R}^{\gamma} \left(\bar w\right)}{2 \pi
\left(\bar z - \bar w\right)}
\nonumber \\
{J}_{L}^{\alpha} \left(z\right) n^{\beta} \left(w,\bar w\right) &\sim&
-\frac{\delta^{\alpha \beta} \epsilon\left(w,\bar w\right)}{4 \pi i
\left(z - w\right)}
- \frac{\epsilon^{\alpha \beta \gamma} n^{\gamma} \left(w,\bar w\right)}{4 \pi i
\left(z - w\right)} 
\nonumber \\
{J}_{R}^{\alpha} \left(\bar z\right) n^{\beta} \left(w,\bar w\right) &\sim&
\frac{\delta^{\alpha \beta} \epsilon\left(w,\bar w\right)}{4 \pi i
\left(\bar z - \bar w\right)}
- \frac{\epsilon^{\alpha \beta \gamma} n^{\gamma} \left(w,\bar w\right)}{4 \pi i
\left(\bar z - \bar w\right)} 
\nonumber \\
n^{ \alpha} \left(z,\bar z\right) {J}_L^{\beta} \left(w\right) &\sim&
\frac{\delta^{\alpha \beta} \epsilon\left(w,\bar w\right)}{4 \pi i
\left(z - w\right)}
- \frac{\epsilon^{ \alpha \beta \gamma} n^{\gamma} \left(w,\bar w\right)}{4 \pi i
\left(z - w\right)}                                      
\nonumber \\
n^{\alpha}  \left(z,\bar z\right) {J}_R^{\beta} \left(\bar w\right) &\sim&
-\frac{\delta^{\alpha \beta} \epsilon\left(w,\bar w\right)}{4 \pi i
\left(\bar z - \bar w\right)}
- \frac{\epsilon^{\alpha \beta \gamma} n^{\gamma} \left(w,\bar w\right)}{4 \pi i
\left(\bar z - \bar w\right)}           
\nonumber \\
 \frac{\pi^2}{\lambda^2} 
n^{\alpha} \left(z,\bar z\right)  n^{ \beta} \left(w,\bar w\right) &\sim&
\frac{\delta^{ \alpha \beta}}{2 a_0 |z-w|}
\nonumber \\
&+&
 \frac{\pi^2\delta^{ \alpha \beta}}{3 a_0 |z-w|}
\left(\left(z-w\right)^2 {\bf J}_L^2\left(w\right) 
\right. \nonumber \\
&+& \left. \left(\bar z - \bar w\right)^2 {\bf J}_R^2\left(\bar w\right) \right)
\nonumber \\
&+& 2 \pi^2 \frac{|z-w|}{a_0} \left(\delta^{\alpha \beta} \left({\bf J}_L 
\cdot {\bf J}_R\right)\left(w,\bar w\right) 
\right. \nonumber \\
&-& \left.
 \left(J_L^{\alpha} J_R^{\beta} + J_L^{\beta} J_R^{\alpha} \right)\left(w,\bar w\right)
\right)
\nonumber \\
&+&  \frac{i \pi \epsilon^{\alpha \beta \gamma}}{a_0 |z-w|} 
\left(\left(z-w\right) J_L^{\gamma} \left(w\right)
\right. \nonumber \\
&+& \left.
\left(\bar z - \bar w\right) J_R^{\gamma} \left(\bar w\right) \right),
\label{opesu2}
\end{eqnarray}
where $z - w = v\tau + i a_0$ and $\bar z - \bar w = v \tau - i a_0$.

With all these results at hand, one can investigate
the leading contribution of the continuum limit 
of the generalized two-leg spin ladder (\ref{laddergen}).
The calculations are very cumbersome and our final result reads as follows:
\begin{eqnarray}
{\cal H}_{\rm gen} &=& \frac{2\pi v}{3} \int dx \; \sum_{a=1}^{N}
\left( {\bf J}_{aL}^2 + {\bf J}_{aR}^2 \right)
\nonumber \\
&+& a_0  \left(J_r - 2 J_d \right)  \sum_{a=1}^{N-1} \int dx \; 
{\bf n}_a \cdot  {\bf n}_{a+1} 
\nonumber \\
&+& \frac{3 a_0 \left(J_{rr} + J_{dd} + 3 J_{\ell\ell} \right)}{\pi^2} \sum_{a=1}^{N-1} 
\int dx \; \epsilon_a \epsilon_{a+1}
\nonumber \\
&+& a_0 \sum_{a=1}^N \int dx \;
{\bf J}_{aL} \cdot {\bf J}_{aR}
\left[ - \gamma + 
2 J_\ell \left( 1 - \lambda^2 \right)
\right.
\nonumber \\
&+& \left. \frac{\lambda^2}{\pi^2} \left[
\lambda^2 \left( J_{dd} + J_{rr} + 3 J_{\ell\ell} \right)
+ J_{rr} -  J_{dd} - 3 J_{\ell\ell} \right]  \right] 
\nonumber \\
&+& a_0 \sum_{a=1}^{N-1}   \int dx \; \left[{\bf J}_{aL} \cdot {\bf J}_{a+1R} 
+ {\bf J}_{aR} \cdot {\bf J}_{a+1L} \right]
\left[J_r + 2 J_d 
\right.
\nonumber \\
&+& \left.
\frac{1+ 4\lambda^4}{2\pi^2} 
\left(J_{dd} - J_{rr} \right) \right],
\label{hamgenladdcont}
\end{eqnarray}
where we have neglected all chiral interactions.

For the spin ladder (\ref{nlegringexchange}) with a ring exchange, we finally get
\begin{eqnarray}
{\cal H}_{\rm ring} &=& \frac{2\pi v}{3} \int dx \; \sum_{a=1}^{N}
\left( {\bf J}_{aL}^2 + {\bf J}_{aR}^2 \right)
\nonumber \\
&+& a_0  J_{\perp}  \sum_{a=1}^{N-1} \int dx \; 
{\bf n}_a \cdot  {\bf n}_{a+1} 
+ \frac{36 a_0 K}{\pi^2} \sum_{a=1}^{N-1} 
\int dx \; \epsilon_a \epsilon_{a+1}
\nonumber \\
&+& a_0 \sum_{a=1}^N \int dx \;
{\bf J}_{aL} \cdot {\bf J}_{aR}
\left[ - \gamma + 
2 K \left( 1 - \lambda^2 \right)
\right.
\nonumber \\
&+& \left. \frac{4 \lambda^2 K}{\pi^2} \left(
3 \lambda^2 
- 1  \right)  \right] 
\nonumber \\
&+& a_0 \sum_{a=1}^{N-1}   \int dx \; \left[{\bf J}_{aL} \cdot {\bf J}_{a+1R} 
+ {\bf J}_{aR} \cdot {\bf J}_{a+1L} \right]
\nonumber \\
&& \left[J_{\perp} + 4 K - 4K \frac{1+ 4\lambda^4}{\pi^2} \right] .
\label{hamringcont}
\end{eqnarray}



%

\end{document}